\RequirePackage{fix-cm}
\documentclass[smallextended]{svjour3}       % onecolumn (second format)
\smartqed  % flush right qed marks, e.g. at end of proof
\usepackage{graphicx}
\immediate\write18{makeindex \jobname.nlo -s nomencl.ist -o \jobname.nls}
\usepackage[T1]{fontenc}
\usepackage[utf8]{inputenc}
\usepackage{natbib}
\usepackage{authblk}
\usepackage{graphicx}
\usepackage[x11names]{xcolor}
\usepackage{graphicx}
\usepackage{subcaption}

\usepackage{amsmath,amssymb,amsfonts}
\usepackage{mathrsfs}

\usepackage[title]{appendix}

\usepackage{balance}
\usepackage{setspace}
\usepackage{etoolbox}

\patchcmd{\thebibliography}{\section*{\refname}}{}{}{}

\makeatletter
\renewcommand\subsection{\@startsection{subsection}{2}{\z@}%
                                     {-3.25ex\@plus -1ex \@minus -.2ex}%
                                     {-1.5ex \@plus .2ex}%
                                     {\normalfont\normalfont\bfseries}}
\renewcommand\subsubsection{\@startsection{subsubsection}{3}{\z@}%
                                     {-3.25ex\@plus -1ex \@minus -.2ex}%
                                     {-1.5ex \@plus .2ex}%
                                     {\normalfont\normalfont\bfseries}}
\usepackage{array}
\usepackage{booktabs}
\usepackage{multirow}
\usepackage{threeparttable}
\usepackage{makecell}
\usepackage{float}
\floatstyle{plaintop}
\restylefloat{table}
\usepackage{dblfloatfix}
\usepackage{caption}

\usepackage{url}

\usepackage{listings}
\usepackage{beramono}

\usepackage{svg}

\usepackage{adjustbox}
\usepackage{afterpage}

\usepackage{algorithm}
\usepackage{algorithmicx}
\usepackage{algpseudocode}

\usepackage{textcomp}
\usepackage{pifont}
\usepackage{qtree}
\usepackage{multicol}
\usepackage{tikz}

\definecolor{lightgray}{rgb}{0.9,0.9,0.9}
\definecolor{darkgray}{rgb}{0.4,0.4,0.4}
\definecolor{purple}{rgb}{0.65,0.12,0.82}
\definecolor{gray}{rgb}{0.4,0.4,0.4}
\definecolor{line-numbers}{rgb}{0.4,0.4,0.4}
\definecolor{tags}{rgb}{1,0,0}
\definecolor{darkblue}{rgb}{0.0,0.0,0.6}
\definecolor{cyan}{rgb}{0.0,0.6,0.6}
\definecolor{mygreen}{rgb}{0.0,0.5,0.0}
\definecolor{myorange}{rgb}{1.0,0.5,0.0}
\definecolor{prolog-bg}{rgb}{0.95,0.95,0.95}
\definecolor{prolog-keyword}{rgb}{0.0,0.0,0.6}
\definecolor{prolog-atom}{rgb}{0.0,0.5,0.0}
\definecolor{prolog-string}{rgb}{0.8,0.1,0.1}

\lstdefinelanguage{openhab}{
  sensitive=true,
  morecomment=[l]//,
  morecomment=[s]{/}{/},
  commentstyle=\color{mygreen}\ttfamily,
  morestring=[b]",
  morestring=[b]',
  stringstyle=\color{purple},
  identifierstyle=\color{darkblue},
  keywordstyle=\color{cyan},
  morekeywords={abstract,any,as,boolean,break,byte,case,catch,char,
    class,const,continue,def,default,do,double,else,extends,false,final,
    finally,float,for,goto,if,implements,import,instanceof,in,int,interface,
    label,long,native,new,null,package,private,protected,public,return,short,
    static,strictfp,super,switch,synchronized,this,throw,throws,transient,
    true,try,void,volatile,while,with,var,val,Member,received,update,changed,
    Item,midnight,noon,is,command,Time,cron,System,started,start,level,Thing,
    Channel,triggered,from,to,newArrayList,as,DecimalType,DateTimeType},
  keywordstyle=[2]\color{myorange},
  morekeywords=[2]{rule,when,then,end}
}

\lstdefinestyle{openhab}{
  language=openhab,
  backgroundcolor=\color{lightgray},
  xleftmargin=0.3cm,
  frame=tlbr,
  framesep=4pt,
  framerule=0pt,
  columns=flexible,
  extendedchars=true,
  basicstyle=\scriptsize\ttfamily,
  showstringspaces=false,
  numbers=left,
  numbersep=4pt,
  tabsize=1,
  breaklines=true,
  showtabs=false,
  frame=leftline,
  literate={\$}{{\textbackslash\$}}1
}

\lstdefinelanguage{Prolog}{
  morekeywords={:-,?,@,=,==,\=,<,>,<=,>=,is},
  keywordstyle=\color{prolog-keyword}\bfseries,
  morecomment=[l]{\%},
  morestring=[b]',
  morestring=[b]",
  sensitive=true,
  commentstyle=\color{darkgray}\itshape,
  stringstyle=\color{prolog-string},
  basicstyle=\ttfamily,
  identifierstyle=\color{prolog-atom},
}

\lstdefinestyle{PrologStyle}{
  language=Prolog,
  backgroundcolor=\color{prolog-bg},
  numbers=left,
  numberstyle=\tiny\color{line-numbers},
  frame=single,
  framesep=4pt,
  framerule=0pt,
  xleftmargin=0.3cm,
  basicstyle=\scriptsize\ttfamily,
  columns=flexible,
  showstringspaces=false,
  breaklines=true,
  tabsize=2,
  literate={\$}{{\textcolor{blue}{\$}}}1
}

\usepackage{hyperref}
\makeatletter
\newenvironment{btHighlight}[1][]
{\begingroup\tikzset{bt@Highlight@par/.style={#1}}\begin{lrbox}{\@tempboxa}}
{\end{lrbox}\bt@HL@box[bt@Highlight@par]{\@tempboxa}\endgroup}

\newcommand\btHL[1][]{%
  \begin{btHighlight}[#1]\bgroup\aftergroup\bt@HL@endenv%
}
\def\bt@HL@endenv{%
  \end{btHighlight}%
  \egroup
}
\newcommand{\bt@HL@box}[2][]{%
  \tikz[#1]{%
    \pgfpathrectangle{\pgfpoint{1pt}{0pt}}{\pgfpoint{\wd #2}{\ht #2}}%
    \pgfusepath{use as bounding box}%
    \node[anchor=base west, fill=orange!30,outer sep=0pt,inner xsep=1pt, inner ysep=0pt, rounded corners=3pt, minimum height=\ht\strutbox+1pt,#1]{\raisebox{1pt}{\strut}\strut\usebox{#2}};
  }%
}
\makeatother

\lstdefinestyle{openhabstyle}{
    language={openhab},
    moredelim=**[is][\btHL]{`}{`}, % Use btHL for highlighting between backticks
    moredelim=**[is][{\btHL[fill=green!30]}]{@}{@}, % Use btHL with green fill for @...@
    moredelim=**[is][{\btHL[fill=cyan!30]}]{~}{~}, % Use btHL with cyan fill for ~...~
    extendedchars=true,
    backgroundcolor=\color{lightgray},
    xleftmargin=0.3cm,
    frame=leftline, % Using leftline frame
    framesep=4pt,
    framerule=0pt,
    columns=flexible,
    basicstyle=\scriptsize\ttfamily,
    showstringspaces=false,
    numbers=left,
    numberstyle=\tiny\color{line-numbers}, % Consistent number style
    numbersep=4pt,
    tabsize=1,
    breaklines=true,
    showtabs=false,
    literate={\$}{{\textbackslash\$}}1,
    label=listing,
    escapeinside={<@}{@>}
}
\definecolor{highlight}{HTML}{FFFFFF}

\usepackage{soul}

\newcommand{\hlc}[2][white]{{%
    \colorlet{foo}{#1}%
    \sethlcolor{foo}\hl{#2}}%
}
\begin{document}

% \title{Cracking IoT Security: Can LLMs Outsmart Static Analysis Tools?}
\title{Cracking IoT Security: Can LLMs Outsmart Static Analysis Tools?\thanks{The data pertaining to this research can be found at \url{https://github.com/JasonQuantrill/llm-v-static-results}}}

\author{Jason Quantrill \and
        Noura Khajehnouri \and
        Zihan Guo  \and
        Manar H. Alalfi %etc.
}

% Removed undefined commands like \orgdiv, \orgname, \city, etc.
\institute{
    Department of Computer Science, Toronto Metropolitan University,\\
    Victoria St, Toronto, M5B 2K3, Ontario, Canada\\
    \email{jason.quantrill@torontomu.ca}\\
    \email{noura.khajehnouri@torontomu.ca}\\
    \email{zihan.guo@torontomu.ca}\\
    \email{manar.alalfi@torontomu.ca (corresponding Author)}\\
}

\maketitle
\abstract{Smart home IoT platforms such as openHAB rely on \emph{Trigger--Action--Condition} (TAC) rules to automate device behavior, but the interplay among these rules can give rise to interaction threats—unintended or unsafe behaviors emerging from implicit dependencies, conflicting triggers, or overlapping conditions. Identifying these threats requires semantic understanding and structural reasoning that traditionally depend on symbolic, constraint-driven static analysis. This work presents the first comprehensive evaluation of Large Language Models (LLMs) across a multi-category interaction threat taxonomy, assessing their performance on both the original openHAB (oHC/IoTB) dataset and a structurally challenging Mutation dataset designed to test robustness under rule transformations.

We benchmark Llama 3.1 8B, Llama 70B, GPT-4o, Gemini-2.5-Pro, and DeepSeek-R1 across zero-, one-, and two-shot settings, comparing their results against oHIT’s manually validated ground truth. Our findings show that while LLMs exhibit promising semantic understanding—particularly on action- and condition-related threats—their accuracy degrades significantly for threats requiring cross-rule structural reasoning, especially under mutated rule forms. Model performance varies widely across threat categories and prompt settings, with no model providing consistent reliability. In contrast, the symbolic reasoning baseline maintains stable detection across both datasets, unaffected by rule rewrites or structural perturbations.

These results underscore that LLMs alone are not yet dependable for safety-critical interaction-threat detection in IoT environments. We discuss the implications for tool design and highlight the potential of hybrid architectures that combine symbolic analysis with LLM-based semantic interpretation to reduce false positives while maintaining structural rigor.}

\keywords{ LLM, IoT, static analysis, safety}

\maketitle
%\newpage
\section{Introduction}\label{sec1}
The rapid adoption of smart home IoT applications has revolutionized home automation, enabling users to create complex, rule-based systems using platforms like openHAB. These systems rely on Trigger-Action-Condition (TAC) frameworks to define automation rules, which, while powerful, can inadvertently introduce Rule Interaction Threats (RITs). RITs occur when the interplay of multiple rules leads to unintended behaviors, such as security vulnerabilities or functional failures. Detecting these threats is challenging, as it requires a deep understanding of both the semantic and structural relationships between rules. Traditional approaches, such as symbolic reasoning-based static analysis tools like oHIT \cite{oHIT_Dataset, oHIT}, have been effective in identifying RITs with high precision. However, these methods often struggle with scalability and adaptability to large, dynamic codebases, limiting their practicality in real-world scenarios.

%The emergence of Large Language Models (LLMs), such as LLaMA and GPT, has opened new avenues for automating security analysis. These models, trained on vast datasets of code and natural language, excel at recognizing patterns and understanding context, making them potentially well-suited for identifying RITs in smart home IoT applications. However, their ability to perform deep semantic reasoning and their effectiveness compared to traditional static analysis tools remain understudied. This gap in research motivates our work, which seeks to evaluate the capability of LLMs in detecting RITs and compare their performance to established symbolic reasoning-based approaches.
%In this paper, we focus on LLaMA 3.1 8B and GPT-4o, a state-of-the-art LLMs, and assess their ability to identify and predict RITs in openHAB rules. Using a dataset of openHAB rules and oHIT’s manually verified outcomes as ground truth. Our study aims to address four key research questions:

The rapid development of Large Language Models (LLMs) has created new possibilities for analyzing security-relevant behaviors in IoT automation systems. Models such as GPT-4o, Llama 3.1 (8B and 70B), Gemini-2.5-Pro, and DeepSeek-R1 are increasingly capable of understanding natural language and code-like rule structures, raising the question of whether they can meaningfully assist in identifying rule-interaction threats (RITs) in platforms like openHAB. While their strengths in pattern recognition and contextual interpretation are well known, their reliability in tasks that require structured reasoning across interacting rules remains uncertain.
In this work, we evaluate a representative set of state-of-the-art LLMs across all experimental conditions, including multi-label RIT classification, fine-grained category differentiation and varying zero-, one-, and two-shot prompting. All models are assessed consistently across both the original oHC/IoTB dataset and a Mutation dataset designed to stress-test robustness. %We examine how the models behave across diverse threat types, where they succeed, and where their reasoning breaks down—particularly in cases involving cross-rule dependencies or transformed rule structures.
Our study is guided by four research questions
\begin{itemize}
\item RQ1 (Baseline Capability): How effective can pre-trained LLMs (standard model) validate and classify Rule Interaction Threats (RITs) in real-world openHAB datasets?, this RQ Evaluates LLM-only performance on real data.
\item RQ2 (Model Scaling Effect): How does the parameter size of LLMs (e.g., Llama 3.1 8B vs 70B) influence their contextual validation accuracy and reasoning consistency when analyzing RITs?, this RQ Explores model scale and reasoning variants.
\item RQ3 (Scalability and Generalizability): Does the approach maintain its performance advantages when applied to large-scale, mutation-based datasets where all instances represent true vulnerabilities? this RQ Tests robustness, scalability, and generalization.
\item RQ4 (Hybrid Effectiveness): To what extent does the proposed hybrid workflow—integrating symbolic analysis with LLM-based contextual validation—improve precision and reduce false positives compared to symbolic-only and LLM-only approaches on real and mutation datasets? this RQ Compares hybrid vs. standalone performance.
\end{itemize}
By answering these questions, we aim to provide critical insights into the potential of LLMs for enhancing security analysis in automation systems. 
\vspace{-0.3cm}
\section{Background}\label{sec2}
OpenHAB rules provide the core logic for home automation by executing instructions in response to system events \cite{openhab}. Each rule is architecturally divided into three fundamental sections:

\begin{itemize}
    \item \textit{Triggers:} These are the starting points or catalysts that initiate a rule's evaluation. A trigger is defined by a specific event, such as a change in an item's state (e.g., a motion sensor switching to ``ON''), an item receiving a command (e.g., a light being turned off), or a temporal event (e.g., a specific time of day or a cron schedule).
    
    \item \textit{Conditions (Optional):} Acting as logical filters, conditions are Boolean expressions that determine whether the rule's actions should proceed. After a trigger fires, the conditions are evaluated; the actions will only execute if all conditions return \textit{true}. For instance, a condition could check that it is after sunset before turning on a light triggered by motion.
    
    \item \textit{Actions:} This section defines the tangible operations performed when a rule is successfully triggered and its conditions are met. Actions are the ``then'' part of the rule, and can range from simple commands (e.g., \textit{Light.send(OFF)}) to complex operations like executing scripts, sending notifications, or performing mathematical calculations on variables.
\end{itemize}

These rules are authored in the Xtend programming language \cite{Xtend}, which offers a more streamlined and expressive syntax compared to standard Java, while maintaining full interoperability with the Java ecosystem. The rule syntax itself is built upon the Xbase and Xtend frameworks, enforcing a structured format that includes a mandatory rule name for identification, a declaration of one or more triggers, an optional set of conditions, and a script block containing the actions. Consequently, an openHAB Rules file (typically with a \textit{.rules} extension) is essentially a modular script composed of multiple, independently operating rules that follow this precise and powerful structure.
%\vspace{-0.3cm}
\begin{lstlisting}[style=openhabstyle]
rule "<RULE_NAME>"
when
  <TRIGGER_CONDITION> [or <TRIGGER_CONDITION2> [or ...]]
then
   <SCRIPT_BLOCK>
end
\end{lstlisting}

In the openHAB ecosystem, automation rules coordinate the behavior of interconnected devices to create responsive, intelligent environments. While this flexibility is a strength, it also introduces opportunities for unintended or even malicious interactions between rules. Such interactions—known as \textit{Rule Interaction Threats (RITs)}—can cause devices to behave unpredictably or unsafely, even when each individual rule is correct and well-intentioned.
Prior work has proposed several taxonomies of interaction threats, often describing similar underlying issues but drawing category boundaries differently or expanding certain behaviors into more granular subclasses. After examining multiple influential studies \cite{TAP}, \cite{fixTAP}, \cite{iruler} we consolidated these perspectives into three foundational categories of rule interaction threats. These categories represent the core logical vulnerabilities arising between rules (i.e., excluding self-interactions), and each can manifest in different forms depending on how rules depend on, trigger, or contradict one another.

\begin{figure}
    \centering
    \includegraphics[width=0.95\linewidth]{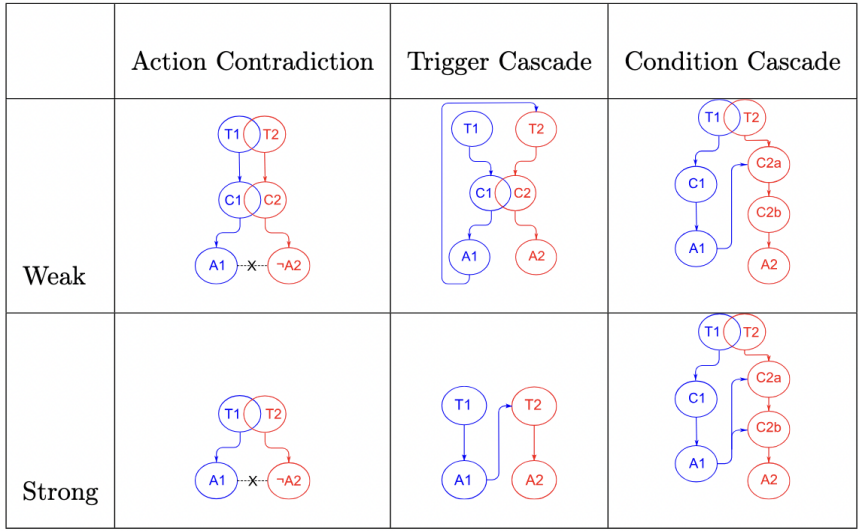}
    \caption{An overview of RITs}
    \label{tab:statemachine}
\end{figure}

Crucially, the presence of an RIT does not imply malicious authorship; many arise simply from complex rule interplay in real-world smart home setups. The following section describes each RIT category in detail and illustrates how these behaviors can surface in practice. A summary of all RIT types is provided in Figure \ref{tab:statemachine}.
%\vspace{0.2cm}
\vspace{-0.3cm}
\subsection{Action Contradiction - AC} Occurs when two rules can occur simultaneously or interrupt one another. These rules have contradicting actions that can lead to undesired behaviour. Action Contradictions are the most common RIT as many rulesets match the overall structure of an AC threat, which are then flagged by oHIT. Due to this, oHIT creators have mentioned detected rules of this nature must be reevaluated by the user to determine if they are truly a threat for the overall automation system.
%\vspace{-0.3cm}
\begin{lstlisting}[style=openhabstyle,caption={AC Example: Rule A}, label={tab:AC1}]  
rule "Turn on Water Sprinkles at Sunset"
when 
    ~Time cron "0 00 18 * * ?"~
then
    `sendCommand(Window_Lock, ON)`
    sendCommand(TurnOnWaterSprinklers, ON)

end
    \end{lstlisting}
%\vspace{-0.2cm}
\begin{lstlisting}[style=openhabstyle,caption={AC Example: Rule B},label={tab:AC2}]  
rule "Open Windows when Temperature is too Hot"
when
    ~Temperature.state >= 25~
then
    `sendCommand(Window_Lock, OFF)`
end
    \end{lstlisting}
%\vspace{-0.1cm}    

An example of an Action Contradiction is illustrated in Listing \ref{tab:AC1} \& \ref{tab:AC2}. 
The \textbf{intended use }of this ruleset is the following: rule 1 is to turn on the water sprinklers every day at sunset, which is 6:00pm and the second rule will open house windows if the temperature inside is higher than 25 degrees Celsius to allow fresh air to enter. However, these two rules can possibly occur at the same time which would result in contradictory results as rule 1 sets Window\_Lock to ON and rule 2 sets Window\_Lock to OFF. 

The \textbf{unintended use} of this rule occurs when rule 1 closes the windows during sunset to turn on the water sprinkler, but we also have hot temperatures inside the house which will cause rule 2 to execute and open the windows. If this unintended behaviour occurs, then the water from the sprinklers can get inside the house and cause damages. There are two different types of Action Contradictions; Weak Action Contradiction and Strong Action Contradiction. In this section, we will explain the differences between both types of AC. 
%%%%%%%%%%%%%%%%%%%%
% Weak Action Contradiction - WAC
%%%%%%%%%%%%%%%%%%%%
\vspace{-0.3cm}
\subsubsection{Weak Action Contradiction - WAC}
%%%%%%%%%%%%%%%%%%%%
% WAC DESCRIPTION
%%%%%%%%%%%%%%%%%%%%
A Weak Action Contradiction occurs when at least one of the rules within the ruleset contains a condition. If both rules have a condition then they must overlap, either in timing or status (item being ON/OFF). The triggers in WAC must also overlap, however, the resulting actions in both rules must be contradictory and as a WAC is a subcategory of an AC, these rules must be able to occur simultaneously.
%%%%%%%%%%%%%%%%%%%%
% Strong Action Contradiction - SAC
%%%%%%%%%%%%%%%%%%%%
\vspace{-0.3cm}
\subsubsection{Strong Action Contradiction - SAC}
%%%%%%%%%%%%%%%%%%%%
% SAC DESCRIPTION
%%%%%%%%%%%%%%%%%%%%
A SAC is similar to a WAC as rulesets of this nature must contain overlapping triggers and contradictory actions, however, there must not be any conditions present in rulesets flagged as a SAC threat.  
%%%%%%%%%%%%%%%%%%%%
% SAC FORM
%%%%%%%%%%%%%%%%%%%%
% \textbf{General Form:} 
% $$\scalemath{0.9}
% {A1=\neg A2 \; \&\& \; T1= T2 \;\&\&\; C1= C2 = \varnothing}$$
\vspace{-0.3cm}
\subsection{Trigger Cascade - TC}
Rulesets flagged as a Trigger Cascade must have a cascading behaviour. As in, one rule’s action must trigger the second rule’s execution. These rulesets can be dangerous as they can lead to a series of unintentional events, which can be unnoticed by users for a long period of time, such as a garage door opening while the user is not home. 
%\vspace{-0.3cm}
    \begin{lstlisting}[style=openhabstyle,caption={TC Example: Rule A},label={tab:TC1}]  
rule "Turn on Lights & other Morning Activities"
when 
    ~Time cron "0 30 08 * * ?" && day.state == "Weekday"~

then
    sendCommand(Kitchen_Light, ON)
    `sendCommand(Foyer_Light, ON)`
    sendCommand(CoffeeMaker, ON)
    sendCommand(MorningNews, ON)

end
    \end{lstlisting}
%\vspace{-0.2cm}    
    \begin{lstlisting}[style=openhabstyle,caption={TC Example: Rule B},label={tab:TC2}]  

rule "Unlock Door & Garage Door"
when
    ~Foyer_Light changed to ON~
then
    @if(time >= 8:00 && time <= 9:00)@
        `sendCommand(Door_Lock, OFF)`
        `sendCommand(Garage_Door, OPEN)`
end
\end{lstlisting}
%\vspace{-0.1cm}  
Listings \ref{tab:TC1} \& \ref{tab:TC2} demonstrate an example of a Trigger Cascade. The \textbf{intended use} of this ruleset is to automate morning activities for the user. Rule 1 turns on activities such as the kitchen and foyer lights, coffee maker and the morning news at a set time (8:30 am on weekdays). In Rule 2, when the foyer light turns on (and if the time is between 8:00 am and 9:00am) then it is assumed the user is getting ready to leave their house for work, and the garage door will open and the front door will lock. When rule 1 sets Foyer\_Light to ON, this will cause rule 2 to execute as its trigger occurs when Foyer\_Light is changed to ON. 

The\textbf{ unintended use} of this ruleset is if the user is not home during the set time or decides not to go work, rule 1 will still execute, which will then trigger rule 2. When this happens, rule 2 will cause the front door and garage door to be left open which makes it easier for intruders to get inside their home. There are two different types of a Trigger Cascade; Weak Trigger Cascade and Strong Trigger Cascade. We will explain the difference between these two types in the following section. 
%%%%%%%%%%%%%%%%%%%%
% Weak Trigger Cascade - WTC
%%%%%%%%%%%%%%%%%%%%
\vspace{-0.3cm}
\subsubsection{Weak Trigger Cascade - WTC}
Similar to a WAC, a Trigger Cascade is defined as weak if at least one of the rules present in the ruleset contains a condition. If both rules contain a condition then they must overlap, meaning that both conditions can occur at the same time. The resulting action of one rule must trigger the execution of the next rule to fit the Trigger Cascade definition. 
% \textbf{General Form:} 
% $$\scalemath{0.9}
% {A1\rightarrow T2 \; \&\& \;  C1\cap C2 \neq \varnothing}$$ 
%%%%%%%%%%%%%%%%%%%%
% Strong Trigger Cascade - STC
%%%%%%%%%%%%%%%%%%%%
\vspace{-0.3cm}
\subsubsection{Strong Trigger Cascade - STC}
%%%%%%%%%%%%%%%%%%%%
% STC DESCRIPTION
%%%%%%%%%%%%%%%%%%%%
This is the simplest RIT. A STC must not contain any conditions, the only requirement that rulesets of this nature must possess is the ability for one rule to trigger the execution of another. A STC is less restrictive than a WTC and it is more likely to lead to catastrophic events due to not containing any additional safeguards (conditions).
%%%%%%%%%%%%%%%%%%%%
% STC FORM
%%%%%%%%%%%%%%%%%%%%
% \textbf{General Form:}  
% $$\scalemath{0.9}
% {A1\rightarrow T2 \; \&\& \;  C1 = \varnothing}$$
\vspace{-0.3cm}
\subsection{Condition Cascade - CC}
A condition cascade requires both rules to at least have one condition where the resulting action in one rule allows the one or all conditions of the second rule to succeed. CC rulesets must also have overlapping triggers, meaning that they could be triggered in parallel. CC threats are less common as it is harder to classify them when compared to AC and TC threats. This is because there are many definition requirements that must be fulfilled in order for a threat to be categorized as a CC.
%\vspace{-0.3cm}
\begin{lstlisting}[style=openhabstyle,caption={CC Example: Rule A},label={tab:CC1}]  
rule "Open Windows when Fire Alarm Rings"
when
    FireAlarmRinging changed to ON
then
    ~if (temperature.state >= 57)~
        `sendCommand(window_Lock, OFF)` 
end
    \end{lstlisting}  
%\vspace{-0.2cm}
    \begin{lstlisting}[style=openhabstyle, caption= {CC Example: Rule B},label={tab:CC2}]
rule "Turn on Air Conditioner at BedTime"
when
    ~Time cron "0 30 22 * * ?"~
then
    `if (window_Lock == OFF)`
        `sendCommand(window_Lock, ON)`
        sendCommand(turnAC, ON)
end
    \end{lstlisting}
%\vspace{-0.1cm}    
Listing \ref{tab:CC1} \& \ref{tab:CC2} demonstrates an example of a Condition Cascade RIT. The intended use of this ruleset is for rule 1 to provide additional safety measures for escaping a fire by unlocking/opening windows and rule 2 to close/lock windows when the AC is set to turn on at 11:30 pm. 

However, \textbf{the unintended use} of this ruleset is when a fire occurs during the time that the AC is set to turn on (bed time). If a fire is detected (temperature of 57 degrees celsius) before bedtime, rule 1 will unlock all the windows to let fresh air into the house and let people escape. However, if bedtime occurs, then rule 2 will lock and close all the windows, which will make it harder for people to escape. Due to the air conditioning and the locked/closed windows, the fire will also spread. These are catastrophic events that can occur, which is why it is important for RITs to be flagged and notified to the user. Similar to AC and TC threats, there are two different variations of CC threats; Weak Condition Cascade and Strong Condition Cascade. We will explain the differences between these two types in the following section. 
%%%%%%%%%%%%%%%%%%%%
% Weak Condition Cascade - WCC
%%%%%%%%%%%%%%%%%%%%
\vspace{-0.3cm}
\subsubsection{Weak Condition Cascade - WCC}
%%%%%%%%%%%%%%%%%%%%
% WCC DESCRIPTION
%%%%%%%%%%%%%%%%%%%%
A WCC must have overlapping triggers, where the resulting action in the first rule must enable one (but not all) conditions guarding an action in the second rule. 
% \textbf{General Form:}  
% $$\scalemath{0.9}
% {A1\rightarrow C2 \space \&\& \space  T1\cap T2 \neq \varnothing}$$
%%%%%%%%%%%%%%%%%%%%
% Strong Condition Cascade - SCC
%%%%%%%%%%%%%%%%%%%%
\vspace{-0.3cm}
\subsubsection{Strong Condition Cascade - SCC}
%%%%%%%%%%%%%%%%%%%%
% SCC DESCRIPTION
%%%%%%%%%%%%%%%%%%%%
A SCC is similar to a WCC, however, action A in the first rule must enable all the conditions that guard action B in the second rule. As a result, a ruleset can still be an SCC even if the second rule only has one condition, as long as the action in the first rule enables that one condition.
\section{Approach}\label{sec3}
\begin{figure}
    \centering
    \includegraphics[width=0.95\linewidth]{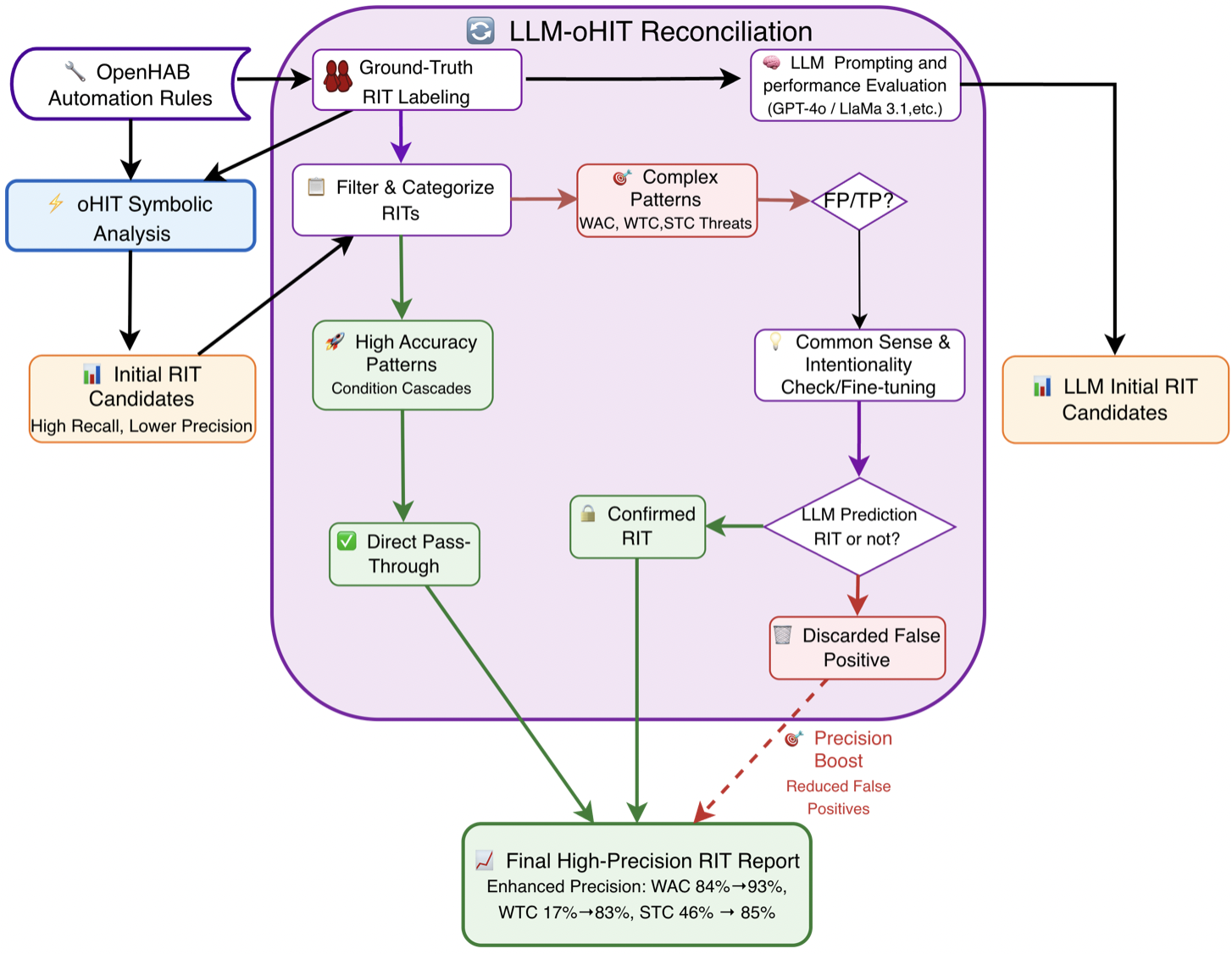}
    \caption{LLM-oHIT Reconciliation Approach }
    \label{fig:Approach}
\end{figure}

Guided by the hybrid workflow’s design presented in Figure \ref{fig:Approach}, our study investigates the interplay between symbolic reasoning and LLM contextual understanding in IoT rule analysis. We first assess the independent performance of LLMs across multiple architectures and parameter scales (RQ1–RQ2). We then evaluate scalability and generalizability, we replicate the experiments on a mutation-based dataset containing exclusively vulnerable cases (RQ3). Finally, we  examine whether integrating these models into a reconciliation-driven hybrid pipeline can reduce false positives and enhance detection precision on real datasets (RQ4) %We then examine whether integrating these models into a reconciliation-driven hybrid pipeline can reduce false positives and enhance detection precision on real datasets (RQ3). Finally, to evaluate scalability and generalizability, we replicate the experiments on a mutation-based dataset containing exclusively vulnerable cases (RQ4).
%Our approach aimed to assess the capability of Large Language Models (LLMs) in identifying interaction threats within openHAB rules, benchmarking their performance against a symbolic reasoning-based static analysis method, oHIT.
\subsection{Workflow Description}
The hybrid detection workflow begins with the Symbolic Analysis phase, where the oHIT tool performs a static analysis of raw openHAB automation rules. Using formal logic and predefined patterns, oHIT systematically scans the code's syntactic and structural properties to generate an initial list of Rule Interaction Threat (RIT) candidates. This approach prioritizes high recall, ensuring most potential threats are flagged, but at the cost of lower precision, as it generates a number of false positives—particularly for complex, context-dependent threats like Weak Action Contradictions (WAC) and Weak Trigger Cascades (WTC).

The process then advances to the core Reconciliation \& Validation phase, where a novel hybrid engine reconciles the outputs of the symbolic and LLM-based analyses. The initial candidate list is first filtered and categorized by threat complexity based on Ground-Truth validation. Unambiguous patterns (e.g., Strong Condition Cascades) that were labeled as TP are fast-tracked via a Direct Pass-Through to the final list, leveraging oHIT's reliability for clear cases and avoiding unnecessary computational overhead. In contrast, complex threats (WAC and WTC) are routed for LLM Contextual Validation. Here, models like GPT-4o are not tasked with primary detection but are instead prompted to validate oHIT's findings through a "Common Sense \& Intentionality Check." The LLM analyzes rule logic, item names, and operational context to determine if an apparent conflict represents a genuine threat or a logical, intended behavior—for instance, distinguishing between a programming error and a designed sequence where one rule turns a light on at sunset and another turns it off at midnight. The LLM's judgment leads to a final reconciliation: threats it validates are promoted to Confirmed RITs, while those it refutes as benign are Discarded, directly reducing the false positive rate.

The workflow culminates in the Final Synthesis phase, where the Direct Pass-Through threats and the LLM-Confirmed RITs are assembled into a High-Precision RIT Report. This final output successfully maintains the high recall of the symbolic tool while incorporating the LLM's understanding to significantly boost overall precision, as demonstrated by empirical improvements where WTC precision rose from 17\% to 83\%.

\subsection{Dataset Composition and Statistics}
Our evaluation uses three datasets comprising 2{,}640 RIT instances (Table~\ref{tab:dataset-stats}). 
The first two---\textbf{openHAB Community (oHC)} and \textbf{IoTBench (IoTB)}---provide 
145 manually verified RIT instances extracted from 38 openHAB rule files. 
\begin{itemize}
    \item \textbf{oHC}: Widely shared, benign rules collected from the openHAB community forums.
    \item \textbf{IoTB}: Rule files from GitHub projects known to contain logical and syntactic issues; these were manually repaired and used as a stress-test corpus.
\end{itemize}
Both datasets were independently reviewed by two researchers to establish the ground-truth threats.

To broaden and balance RIT coverage, we also used a \textbf{synthetically mutated dataset} generated by our rule-interaction mutator framework~\cite{oHITMutaion}. 
This framework integrates a Python input processor with a TXL-based mutation engine. 
The input processor orchestrates preprocessing, identifies rule pairs eligible for each mutational operator, applies user-controlled or exhaustive selection strategies.

The TXL engine performs grammar-aware source-to-source transformations using a dedicated TXL grammar extension built atop the Java Language Specification (Edition~15) to support openHAB's Xbase/Xtend syntax. Each mutational operator (WAC, SAC, WTC, STC, WCC, SCC) is implemented as a TXL rule that rewrites the selected rule pair according to its structural preconditions, injecting a specific interaction threat while preserving syntactic validity.

Using this framework, we mutated 15 benign openHAB rulesets. For each ruleset, the engine enumerated all valid Rule~A--Rule~B pairs for each operator and injected exactly one covert interaction per mutated ruleset, producing \textbf{2{,}495 unique rulesets}. The resulting mutants provide high-fidelity, semantically meaningful RIT instances spanning the eight foundational categories of rule-interaction threats.
Together, these three datasets form the basis for evaluating oHIT across a diverse and representative range of interaction threats.

\hlc[highlight]{The distribution across datasets reveals important characteristics. In the combined real-world datasets (oHC + IoTB), Action Contradictions (WAC: 85, SAC: 42) represent 87.6\% of all threats, reflecting their natural prevalence in openHAB deployments. Trigger Cascades (WTC: 2, STC: 6) and Condition Cascades (WCC: 4, WCC: 6) are significantly underrepresented, with fewer than 10 instances each.}

\hlc[highlight]{The Mutation dataset addresses this imbalance by providing substantially more examples of all threat types, particularly the rare Condition Cascades (SCC: 723, WCC: 623) and Trigger Cascades (WTC: 298, STC: 174). This balanced representation enables more robust evaluation of model performance across all RIT categories. Initial experiments} (Tables~\ref{table:experiment-a}--\ref{table:experiment-d}) \hlc[highlight]{use the combined oHC and IoTB datasets (145 instances). Expanded evaluation using the Mutation dataset is presented in }Section\ref{sec-mutation} (Tables~\ref{tab:experiment_a_mutation}--\ref{tab:experiment_d_mutation})

\begin{table}[t!]
\centering
\begin{adjustbox}{width=\textwidth}
\begin{tabular}{lcccccccc}
\toprule
\textbf{Dataset} & \textbf{Files} & \textbf{Rules} & \textbf{WAC} & \textbf{SAC} & \textbf{WTC} & \textbf{STC} & \textbf{WCC} & \textbf{SCC} \\
\midrule
oHC      & 12 & 54  & 19  & 8   & 0   & 0   & 0   & 2   \\
IoTB     & 26 & 102 & 66  & 34  & 2   & 6   & 4   & 4   \\
Mutation & --  & --  & 301 & 376 & 298 & 174 & 623 & 723 \\
\midrule
\textbf{Total} & \textbf{38+} & \textbf{156+} & \textbf{386} & \textbf{418} & \textbf{300} & \textbf{180} & \textbf{627} & \textbf{729} \\
\bottomrule
\end{tabular}
\end{adjustbox}
\caption{\small{Ground truth dataset composition showing the distribution of Rule Interaction Threats. The oHC and IoTB datasets consist of manually verified real-world openHAB rules, while the Mutation dataset contains synthetically generated threats (generated from the original IoTB and oHC datasets). Each RIT represents a specific pair of interacting rules. The Mutation dataset statistics (files and rules) are marked as unavailable (--) pending completion of the experimental evaluation.}}
\label{tab:dataset-stats}
\end{table}
An example of the results for a WAC interaction threat is illustrated in Listing \ref{tab:tool-output}.
\vspace{0.3cm}
\begin{lstlisting}[caption={Example of oHIT Output},label={tab:tool-output}]
FILE: detect-output\oh-rules\WateringSystem.rules
------------------------------------------------
THREATS DETECTED: 23
SAC: 0
WAC: 18
STC: 1
WTC: 4
SCC: 0
WCC: 0
------------------------------------------------

1. WAC THREAT DETECTED
    THREAT PAIR: (r1a1, r7a7)
    
    RULES:
        RULE_A [r1]: ("1 Watering_garden_startup")
        RULE_B [r7]: ("7 Watering_starting/stoping")

    OVERLAPPING TRIGGERS:
        TRIGGERS_A:     [r1t1]: System started 
        
        TRIGGERS_B:     [r7t1]: Item notification_proxy_wtr received update

    OVERLAPPING CONDITIONS:
        CONDITIONS_A:   [c0]:   no conditions guarding action 
        22	
        CONDITIONS_B:   [r7c8]: if (msg == "START")
                    AND [r7c9]: if (wtrfronttime > 0)

    CONTRADICTORY ACTIONS:
        ACTION_A:       [r1a1]: wtrvalvefront.sendCommand(off_r)
        ACTION_B:       [r7a7]: wtrvalvefront.sendCommand(on_r)

    THREAT DESCRIPTION: 
    IF OVERLAPPING SETS OF TRIGGERS AND CONDITIONS ARE CONCURRENTLY ACTIVATED
    CONTRADICTORY ACTION EXECUTION COULD OCCUR IN ANY ORDER
    WHICH MAY RESULT IN AN INDETERMINATE DEVICE STATE.

\end{lstlisting} 
Listing \ref{tab:tool-output} displays the analyzed file, the number of RITs identified, and the reasoning for classifying the ruleset as a WAC. The output specifies the rule names, overlapping triggers and conditions, and contradictory actions. This ruleset is classified as a WAC because the triggers could overlap—the system might restart while the watering process is starting or stopping. If this occurs, the water valve could turn off when it should remain on, as indicated by the red highlighted text.

\subsection{Model Selection and Experimental Setup}
We employ two categories of models—\textit{standard Large Language Models (LLMs)} and \textit{Large Reasoning Models (LRMs)}—to evaluate their effectiveness in validating and classifying Rule Interaction Threats (RITs). All models are used consistently across all experiments to enable a fair comparison of baseline capability, scaling effects, reasoning consistency, and scalability.

All models were evaluated using a two-tier hardware configuration separating local execution from large-scale server-side inference. Local experiments were conducted on a workstation equipped with an NVIDIA RTX 3070 Ti Laptop GPU (8 GB VRAM) running NVIDIA Driver 581.29. Large-parameter models, including Llama 3.1 70B, were executed on a server cluster with 8× NVIDIA H100 GPUs, NVIDIA Studio Driver 581.42, and CUDA Toolkit 13.0. Local inference was managed through Ollama v0.12.10.

All open-source models—Llama 3.1 (8B and 70B) and DeepSeek-R1 7B—were quantized to q4k\_m, resulting in memory footprints of approximately 6.3 GB, 45 GB, and 5.3 GB, respectively. Local executions used Ollama’s default decoding configuration (temperature = 0.8, top\_p = 0.9, top\_k = 40) \cite{ollama_docs}. Proprietary baselines employed the Nov-2024 GPT-4o snapshot \cite{openai_gpt4o}and Gemini-2.5-Pro (June 2025) \cite{google_gemini}.

This environment supported consistent evaluation across both standard LLMs and reasoning-oriented LRMs while enabling controlled comparison between edge-level and cloud-based inference.

Together, these five models provide a balanced experimental framework that spans open- and closed-source systems, small- and large-scale architectures, and standard versus reasoning-enhanced capabilities—enabling a comprehensive evaluation of performance, reasoning consistency, and scalability across all experimental settings.

\subsection{Prompting Strategy}
%We compare the performance of two LLMs: an open-source model, LlaMa 3.1 8b, and a closed-source model, GPT-4o. 
Our experiments employ prompt-based methods, which are critical for guiding the models’ predictions and ensuring responses align with the desired output criteria. These prompts include task descriptions, background information in natural language, and output format specifications. This approach leverages the generalized pre-trained capabilities of the LLMs without requiring task-specific fine-tuning.

Each prompt begins by defining the task: identifying RITs in openHAB automation rules. It then explains the rule structure, highlighting key components such as rule triggers, which are located between the “when” and “then” keywords. Finally, the prompt specifies the output format, requiring the model to return one or more RITs in the format: WAC or STC,WAC (if multiple answers are returned). The basic structure of the prompts is illustrated in Listing \ref{tab:prompt-example}.

\begin{lstlisting}[caption={The structure of the prompts},label={tab:prompt-example}]
You are an expert system for detecting Rule Interaction Threats (RITs) in openHAB automation rules. Your task is to analyze rules and identify specific threat patterns.

RULE STRUCTURE
Rules follow this format:
rule "Name"
when
    [TRIGGER]
then
    [CONDITIONS & ACTIONS]
end

Key Components:
- Triggers: Found between "when" and "then"
- Actions: Commands using sendCommand or postUpdate
- Conditions: If-statements guarding actions, variables CHANGED to a value 
".state" after a variable simply checks the value of the variable 

THREAT TYPES AND PATTERNS
[Explanations of WAC, SAC, WTC, etc.based of the type of prompting] 

OUTPUT FORMAT
Return only the 3-letter acronyms of detected threats, separated by commas if multiple exist. Do not give me an explanation or any reasoning. 
Do not give me any other output besides the 3 letter acronym.
Example: WAC,STC
Think about your answer before responding. Find the best analysis of the rules. The order of the rules does not matter.
The rules that you must analyze are:
\end{lstlisting} 

The experiments were conducted using three prompt-based methods:

\textbf{Zero-Shot Prompting}: is the simplest prompting method. In this method, the prompt in Listing \ref{tab:zeroshot-prompt} is used, with RITs described solely in natural language and without any examples. The prompt is concatenated with a ruleset from the dataset and input into the LLM. This method is termed "zero-shot" because it relies entirely on the descriptions in the prompt to guide the LLM in generating the correct response, without the use of additional examples to instruct the model.

\begin{lstlisting}[caption={Description of WAC/SAC RITs included in prompt for Zero-Shot Prompting},label={tab:zeroshot-prompt}]
1. ACTION CONTRADICTIONS
Definition: Two rules with conflicting actions and overlapping triggers that could create race conditions.

A. Weak Action Contradiction (WAC)
Features:
- Overlapping triggers
- Overlapping conditions (at least one action guarded)
- Contradictory actions


B. Strong Action Contradiction (SAC)
Features:
- Overlapping triggers
- No guarding conditions
- Contradictory actions
\end{lstlisting}

\textbf{One-Shot Prompting}: In this method, the prompt is modified to include a description of each rule interaction threat (RIT) along with one example of each threat. By incorporating examples into the prompt, the model’s performance is expected to improve, as it receives additional guidance on the patterns to identify for each ruleset. Listing \ref{tab:oneshot-prompt} demonstrates how these examples are integrated into the prompts.

\begin{lstlisting}[caption={ Example of One-Shot Prompting for WAC RIT}, label={tab:oneshot-prompt}]
1. ACTION CONTRADICTIONS
Definition: Two rules with conflicting actions and overlapping triggers that could create race conditions.

A. Weak Action Contradiction (WAC)
Features:
- Overlapping triggers
- Overlapping conditions (at least one action guarded)
- Contradictory actions

Example:
rule "Carbon Monoxide Alert"
when
    Item CO_Detector changes to ON
then
    sendCommand(Windows, OPEN)
    sendCommand(Fans, ON)
end

rule "Potential Security Threat Detected"
when
    Item OutdoorSecurityCamera_Motion changed to ON
then
    if (Security_Mode == ON) {
        sendCommand(Windows, CLOSE)
        sendCommand(DoorLock, ON)
    }
end

Risk: If CO is detected and motion triggers simultaneously, windows may close when they need to stay open.

\end{lstlisting}

\textbf{Two-Shot Prompting:} This method is similar to one-shot prompting but extends it by including two examples of each rule interaction threat (RIT) within the prompt. Providing additional examples may enhance the model’s performance, as it offers more diverse contexts for the model to learn from and identify patterns.

\section{Evaluation Protocol, Metrics, and Scoring Methodology}
This study evaluates the capability of large language models (LLMs) to classify Rule Interaction Threats (RITs) by leveraging two complementary datasets and a metric design that accounts for the substantial class imbalance inherent in RIT distributions. Our methodology isolates the classification problem to determine how well LLMs can differentiate between distinct RIT categories, rather than detect threats outright.

For each RIT instance in our dataset, we extracted the pair of interacting rules and presented them to the LLM with a prompt asking it to classify the type of interaction threat. 
We utilize results from a static analysis benchmark applied to a dataset of rules for various openHAB applications. The benchmark, oHIT\cite{oHIT_Dataset}, analyzes all rule combinations within each file and identifies all categories of RITs. By using the same dataset and the benchmark’s manually verified outcomes as ground truth, we can evaluate the performance of a pre-trained LLM in classifying RITs and compare its results to those of the static analysis tool. 

Our evaluation focuses on threat classification rather than threat detection. For each experiment, we evaluate the LLM's ability to correctly classify rule pairs that are known to contain at least one RIT. We extracted individual RIT instances from the ground truth dataset, where each instance consists of a pair of interacting rules and its corresponding RIT classification. This experimental design tests the LLM's understanding of rule interactions and its ability to distinguish between different threat types. As such, in the LLM-only experiments, we measure only recall, since we are feeding it only known threats (ie false positives are not possible, and precision cannot be usefully calculated) to test whether it can classify them correctly. To calculate recall we use the following formula:
$$ \text{Recall} = \frac{\text{TP}}{\text{TP} + \text{FN}} $$

\subsection*{Evaluation Metrics and Computation Process}
To accurately assess model performance across the six individual Rule Interaction Threat (RIT) categories (WAC, SAC, WTC, STC, WCC, SCC) and the three aggregated categories (AC, TC, CC), we adopted a metric strategy that accounts for the substantial class imbalance present in both the real-world and mutation datasets. The distributions in the 6-class mutation dataset---301 (WAC), 376 (SAC), 298 (WTC), 174 (STC), 623 (WCC), and 723 (SCC)---and in the 3-class dataset---677 (AC), 472 (TC), and 1346 (CC)---demonstrate that directly averaging per-class results would yield misleading conclusions. Accordingly, our evaluation distinguishes between \textit{overall accuracy} and \textit{per-class recall}, ensuring that both aggregate performance and category-specific detection capability are reported clearly and consistently.
\subsection*{Overall Accuracy (Micro Accuracy)}
The \textit{Total Correct} column reported in all result tables corresponds to \textbf{overall (micro) accuracy}, computed as the ratio of all correctly predicted samples to the total number of samples in the dataset:
\[
\text{Overall Accuracy} = \frac{\text{Total Correct Predictions}}{\text{Total Number of Samples}}.
\]
This metric reflects performance proportionally to the true class distribution and is therefore sensitive to the underlying imbalance. For example, in Experiment~C (zero-shot), Table \ref{tab:experiment_c_mutation}, Gemini~2.5~Pro correctly predicted 2,188 out of 2,495 samples, yielding an overall accuracy of 87.70\%. Importantly, this score is not derived by averaging category-level percentages; instead, it represents the model's correctness over the entire dataset.
\subsection*{Per-Class Recall}
The individual RIT category columns (e.g., AC, TC, CC) represent \textbf{per-class recall}, defined as:
\[
\text{Recall}_i = \frac{\text{Correct Predictions for Class } i}{\text{Total True Samples of Class } i}.
\]
These values quantify how effectively the model identifies instances \emph{within} each class, independent of how frequently that class appears. For instance, in the same experiment, the per-class recalls for Gemini~2.5~Pro---59.53\% for AC, 94.28\% for TC, and 99.55\% for CC---reflect the model's class-specific detection ability. All percentages were computed to two significant figures using a consistent rounding approach across all experiments.

\subsection*{Rationale for Metric Selection}
The combination of \textbf{overall accuracy} and \textbf{per-class recall} provides a balanced and interpretable view of model performance. Overall accuracy aligns with real-world deployment scenarios, where correct classification rates depend on naturally skewed distributions. Meanwhile, per-class recall prevents majority classes from overshadowing performance weaknesses in minority categories. Together, these metrics offer a comprehensive representation of both aggregate model behavior and fine-grained category-level effectiveness.
\subsection*{Evaluation Conditions}
We conducted experiments under two distinct conditions to explore different deployment scenarios:
\hlc[highlight]{\textbf{Multiple Responses Allowed (Experiments A, C, E):}
In this condition, the LLM may output multiple RIT classifications for a single rule pair (e.g., "WAC, STC"). We consider the prediction correct if any of the predicted RIT types matches the ground truth label. This approach simulates a deployment scenario where the system flags multiple potential threats for manual review, maximizing recall at the potential cost of precision.}  \hlc[highlight]{\textbf{Single Response Only (Experiments B, D, F):}
In this condition, the LLM is constrained to output exactly one RIT classification. Predictions are scored as correct only if the single output exactly matches the ground truth label. This stricter evaluation provides a clearer measure of classification precision and simulates a deployment scenario where the system must definitively classify each detected threat without ambiguity.}
\hlc[highlight]{The contrast between these two conditions reveals important trade-offs: multiple responses maximize coverage but introduce noise, while single responses demand precision but may miss valid threats when the model is uncertain.}
\section{Results and Discussion}
To address \textbf{RQ1 (Baseline Capability)}—which examines how effectively pre-trained LLMs can validate and classify Rule Interaction Threats (RITs) in real-world \textit{openHAB} datasets—the results highlight a clear distinction in baseline reasoning performance across contradiction categories. Among the evaluated models, \textit{Llama-8b} demonstrates the strongest zero-shot capability, achieving a total accuracy of 64.83\% without any contextual examples. Its high accuracy in \textit{Weak Action Contradiction (WAC)} (80.49\%) and \textit{Weak Condition Contradiction (WCC)} (75\%) suggests that smaller, pre-trained LLMs can reliably detect explicit or surface-level inconsistencies in rule logic. However, its poor results in \textit{Strong Action Contradiction (SAC)} (51.11\%), \textit{Strong Trigger Contradiction (STC)} (16.67\%), and \textit{Strong Condition Contradiction (SCC)} (0\%) reveal clear limitations in resolving complex or multi-layered contradictions that require deeper semantic reasoning. Moreover, the degradation in accuracy under one-shot (36.55\%) and two-shot (39.31\%) settings indicates that contextual examples may not enhance, and can even impair, baseline reasoning for smaller models—likely due to rigid contextual encoding and limited adaptability in few-shot settings.
\begin{table}[t!]
\centering
\begin{adjustbox}{width=\textwidth}
\begin{tabular}{l|ccccccc}
\toprule
\textbf{Model} & \textbf{WAC} & \textbf{SAC} & \textbf{WTC} & \textbf{STC} & \textbf{WCC} & \textbf{SCC} & \textbf{Total Correct} \\
\midrule
Llama-8b Zero-Shot & 80.49\% & 51.11\% & 50.00\% & 16.67\% & 75.00\% & 0.00\% & 64.83\% \\
Llama-8b One-Shot & 53.66\% & 15.56\% & 0.00\% & 0.00\% & 50.00\% & 0.00\% & 36.55\% \\
Llama-8b Two-Shot & 54.88\% & 15.56\% & 100.00\% & 33.33\% & 25.00\% & 0.00\% & 39.31\% \\
\midrule
Llama-70b Zero-Shot & 67.07\% & 31.11\% & 0.00\% & 33.33\% & 50.00\% & 0.00\% & 50.34\% \\
Llama-70b One-Shot & 52.44\% & 2.22\% & 0.00\% & 16.67\% & 25.00\% & 0.00\% & 31.72\% \\
Llama-70b Two-Shot & 51.22\% & 2.22\% & 50.00\% & 16.67\% & 25.00\% & 0.00\% & 31.72\% \\
\midrule
GPT-4o Zero-Shot & 32.93\% & 60.00\% & 100.00\% & 100.00\% & 0.00\% & 66.67\% & 45.52\% \\
GPT-4o One-Shot & 25.61\% & 33.33\% & 50.00\% & 100.00\% & 50.00\% & 66.67\% & 54.27\% \\
GPT-4o Two-Shot & 29.27\% & 40.00\% & 50.00\% & 100.00\% & 75.00\% & 66.67\% & 60.16\% \\
\midrule
Gemini-2.5-Pro Zero-Shot & 75.61\% & 48.89\% & 50.00\% & 100.00\% & 25.00\% & 50.00\% & 65.52\% \\
Gemini-2.5-Pro One-Shot & 43.90\% & 53.33\% & 50.00\% & 100.00\% & 25.00\% & 66.67\% & 49.66\% \\
Gemini-2.5-Pro Two-Shot & 47.56\% & 35.56\% & 0.00\% & 100.00\% & 25.00\% & 50.00\% & 44.83\% \\
\midrule
Deepseek-r1-7b Zero-Shot & 56.10\% & 53.33\% & 50.00\% & 33.33\% & 25.00\% & 16.67\% & 51.72\% \\
Deepseek-r1-7b One-Shot & 75.61\% & 26.67\% & 0.00\% & 16.67\% & 25.00\% & 16.67\% & 53.10\% \\
Deepseek-r1-7b Two-Shot & 64.63\% & 6.67\% & 50.00\% & 0.00\% & 0.00\% & 0.00\% & 39.31\% \\
\bottomrule
\end{tabular}
\end{adjustbox}
\caption{Experiment A: Using all 6 categories of RITs and allowing multiple responses}
\label{table:experiment-a}
\end{table}
In contrast, \textit{Llama-70b}, despite its significantly larger parameter size, exhibits inconsistent performance, achieving 50.34\% in the zero-shot setup but dropping to 31.72\% in both one-shot and two-shot configurations. This instability suggests that model scale alone does not guarantee improved contextual reasoning. The results imply that larger models without sufficient alignment or reasoning optimization may overfit to local context cues, leading to degraded performance when tasked with validating RITs that involve subtle cross-rule dependencies.

Addressing \textbf{RQ2 (Model Scaling Effect)}—which investigates how model size influences reasoning consistency and contextual validation accuracy—the findings reveal that \textit{GPT-4o} outperforms both Llama models across all settings, showing a strong in-context learning trajectory. Its total accuracy increases steadily from 45.52\% (zero-shot) to 54.27\% (one-shot) and 60.16\% (two-shot), demonstrating effective adaptation when provided with minimal examples. Unlike the Llama models, \textit{GPT-4o} sustains perfect accuracy (100\%) in both \textit{Weak Trigger Contradiction (WTC)} and \textit{Strong Trigger Contradiction (STC)} across all configurations and achieves consistently high scores in \textit{Strong Condition Contradictions (SCC)} (66.67\%), indicating advanced contextual integration and multi-step reasoning.

\begin{figure}[t!]
    \centering
    \begin{subfigure}[b]{0.45\textwidth}
        \centering
        \includegraphics[width=\textwidth]{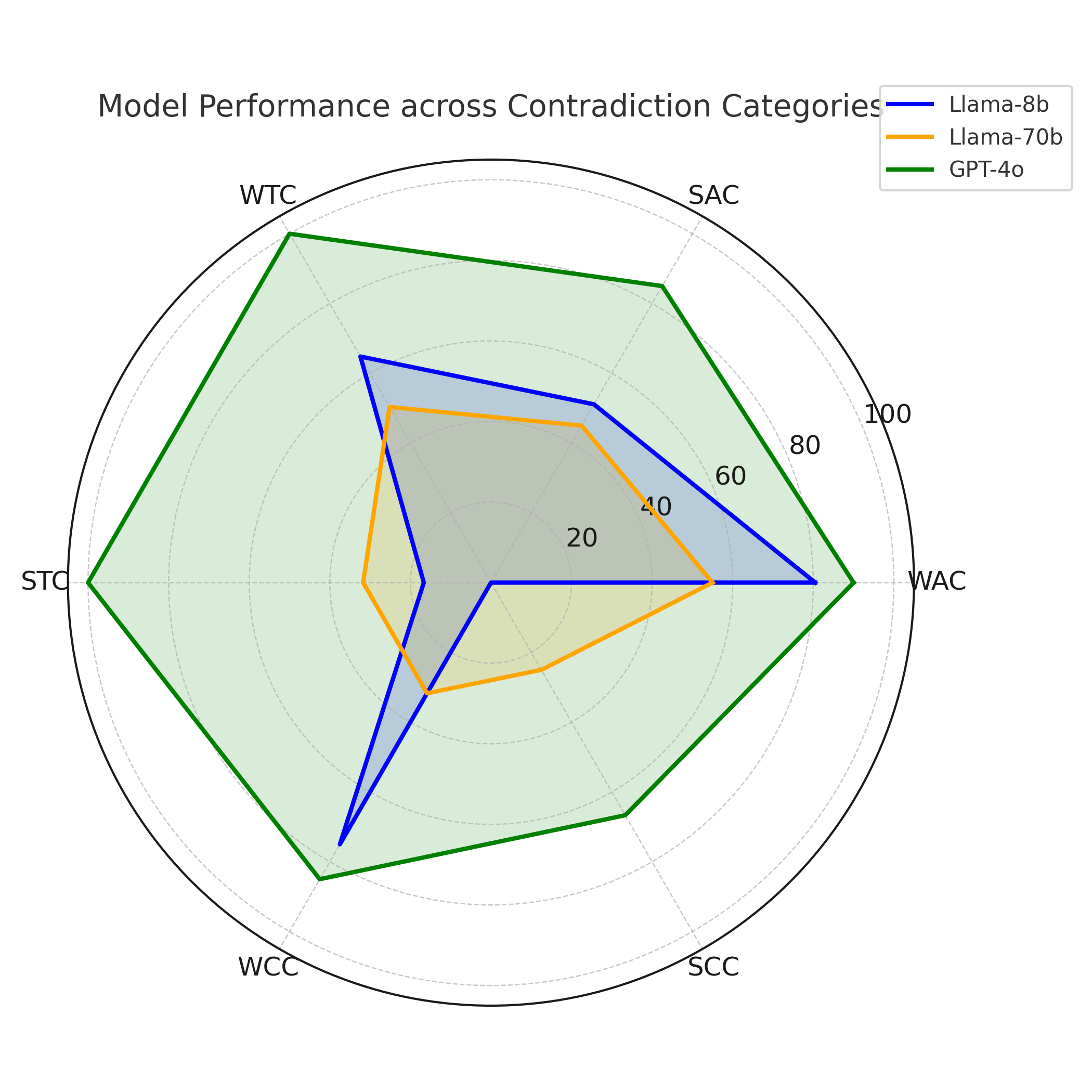}
        \caption{Model performance across contradiction categories (WAC, SAC, WTC, STC, WCC, SCC).}
        \label{fig:radar}
    \end{subfigure}
    \hfill
    \begin{subfigure}[b]{0.48\textwidth}
        \centering
        \includegraphics[width=\textwidth]{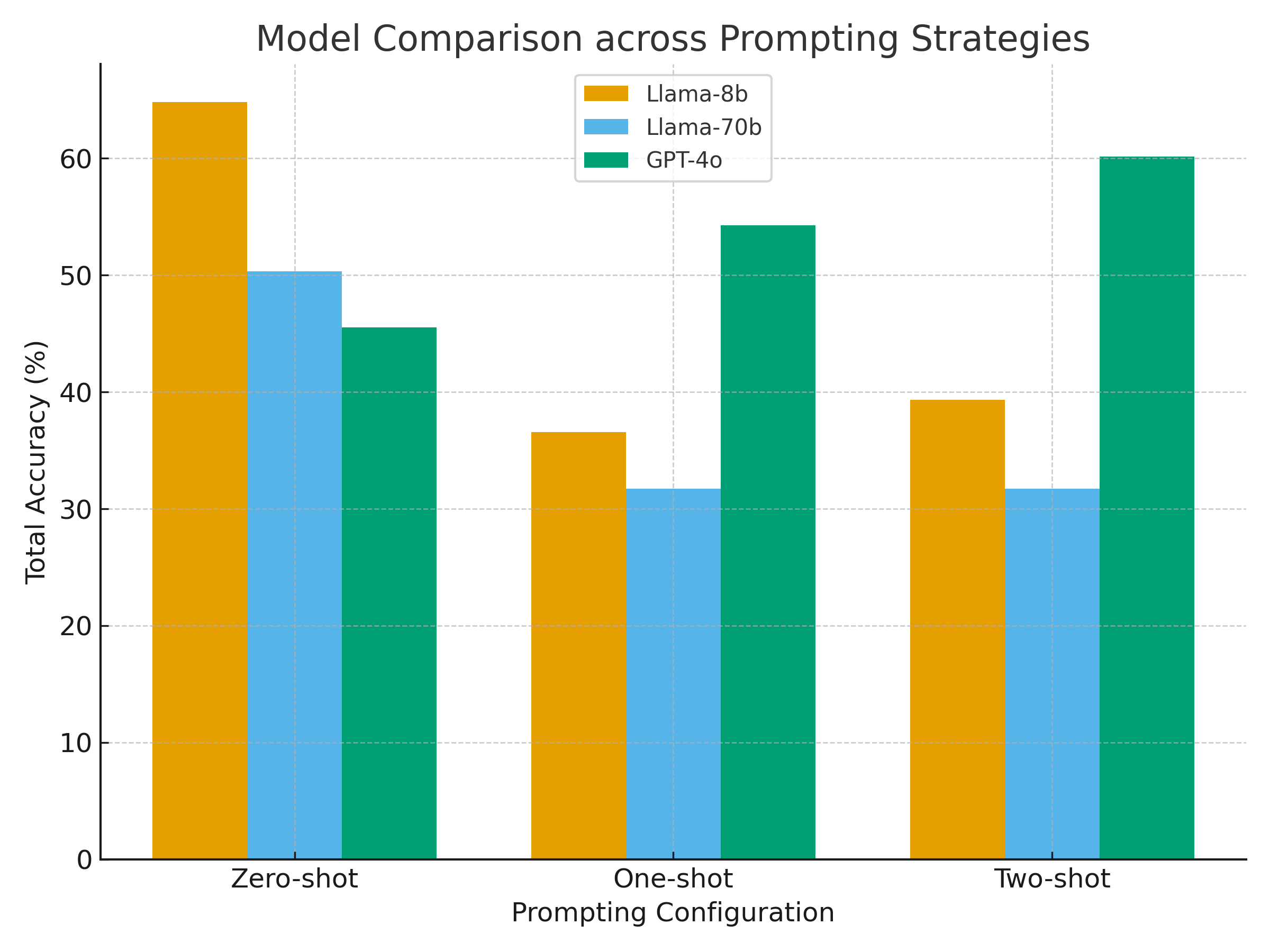}
        \caption{Comparison of total accuracy trends across prompting configurations.}
        \label{fig:barplot}
    \end{subfigure}
    \caption{Comparison of LLM reasoning performance and scaling behavior across RIT contradiction types and prompting configurations.}
    \label{fig:combined}
\end{figure}

The radar visualization (Figure~\ref{fig:radar}) reinforces this interpretation: GPT-4o maintains a broad and balanced performance profile across all contradiction types, while both Llama variants show sharp asymmetries, excelling only in weak contradictions but collapsing in strong ones. The bar plot comparison (Figure~\ref{fig:barplot}) further illustrates this contrast—GPT-4o exhibits steady performance improvement with increasing contextual information, whereas the Llama models decline. Collectively, these findings confirm that scaling and reasoning alignment jointly determine model robustness: parameter size alone is insufficient, but when coupled with effective reasoning optimization—as in GPT-4o—it enables consistent and contextually aware validation of RITs.

\begin{table}[t!]
\centering
\begin{adjustbox}{width=\textwidth}
\begin{tabular}{l|ccccccc}
\toprule
\textbf{Model} & \textbf{WAC} & \textbf{SAC} & \textbf{WTC} & \textbf{STC} & \textbf{WCC} & \textbf{SCC} & \textbf{Total Correct} \\
\midrule
Llama-8b Zero-Shot & 18.29\% & 0.00\% & 0.00\% & 16.67\% & 75.00\% & 0.00\% & 13.10\% \\
Llama-8b One-Shot & 4.88\% & 2.22\% & 0.00\% & 16.67\% & 25.00\% & 0.00\% & 4.83\% \\
Llama-8b Two-Shot & 10.98\% & 0.00\% & 50.00\% & 0.00\% & 25.00\% & 0.00\% & 7.59\% \\
\midrule
Llama-70b Zero-Shot & 51.22\% & 0.00\% & 0.00\% & 50.00\% & 50.00\% & 0.00\% & 32.41\% \\
Llama-70b One-Shot & 50.00\% & 0.00\% & 50.00\% & 33.33\% & 0.00\% & 0.00\% & 30.34\% \\
Llama-70b Two-Shot & 28.05\% & 13.33\% & 100.00\% & 50.00\% & 0.00\% & 0.00\% & 23.45\% \\
\midrule
GPT-4o Zero-Shot & 29.27\% & 53.33\% & 100.00\% & 100.00\% & 25.00\% & 50.00\% & 59.60\% \\
GPT-4o One-Shot & 25.61\% & 31.11\% & 50.00\% & 100.00\% & 100.00\% & 66.67\% & 62.23\% \\
GPT-4o Two-Shot & 13.41\% & 28.89\% & 0.00\% & 83.33\% & 75.00\% & 100.00\% & 50.11\% \\
\midrule
Gemini-2.5-Pro Zero-Shot & 51.22\% & 51.11\% & 50.00\% & 100.00\% & 25.00\% & 50.00\% & 52.41\% \\
Gemini-2.5-Pro One-Shot & 47.56\% & 51.11\% & 100.00\% & 100.00\% & 25.00\% & 50.00\% & 51.03\% \\
Gemini-2.5-Pro Two-Shot & 46.34\% & 53.33\% & 0.00\% & 100.00\% & 25.00\% & 50.00\% & 49.66\% \\
\midrule
Deepseek-r1-7b Zero-Shot & 3.66\% & 6.67\% & 0.00\% & 66.67\% & 0.00\% & 0.00\% & 6.90\% \\
Deepseek-r1-7b One-Shot & 4.88\% & 0.00\% & 0.00\% & 66.67\% & 0.00\% & 0.00\% & 5.52\% \\
Deepseek-r1-7b Two-Shot & 4.88\% & 2.22\% & 0.00\% & 33.33\% & 0.00\% & 0.00\% & 4.83\% \\
\bottomrule
\end{tabular}
\end{adjustbox}
\caption{Experiment B: Using all 6 categories of RITs and NOT allowing multiple responses}
\label{table:experiment-b}
\end{table}
\subsection*{Impact of Single-Response Constraint:}
When models were constrained to output a single RIT classification (Experiment~B), overall accuracy dropped significantly across all architectures compared to the multiple-response setup. This decline highlights the precision--recall trade-off inherent in LLM-based threat validation. Under the stricter single-response condition, smaller models such as \textit{Llama-8b} and \textit{Deepseek-r1-7b} experienced substantial reductions in total accuracy---falling below 15\% and 7\%, respectively---indicating limited confidence and generalization when forced to commit to one interpretation. \textit{Llama-70b} maintained modest stability (approximately 30\%) but still failed to leverage its scale advantage effectively. In contrast, reasoning-optimized models such as \textit{GPT-4o} and \textit{Gemini-2.5-Pro} demonstrated relative resilience, achieving over 50\% accuracy even under single-label constraints. \textit{GPT-4o}, in particular, preserved robust performance (59.6--62.2\%), suggesting superior confidence calibration and internal reasoning consistency.

Comparing these findings with the multiple-response results reveals that allowing multiple predictions inflated recall but masked uncertainty, while the single-response condition exposes each model’s true precision and reasoning discipline. The contrast confirms that while larger, alignment-optimized models sustain balanced reasoning under stricter constraints, smaller or less aligned ones tend to overfit contextual cues and struggle with definitive classification. Thus, the transition from multi- to single-response evaluation provides a clearer indicator of model reliability for deployment in autonomous RIT validation systems.

In summary, under RQ1, pre-trained LLMs demonstrate limited yet nontrivial capability to identify RITs, particularly in weak contradiction contexts. However, when subjected to the single-response constraint, these capabilities diminish significantly across smaller models, revealing challenges in confidence calibration and decisive reasoning when ambiguity is removed. Under RQ2, scaling effects remain significant only when accompanied by reasoning alignment---highlighting that larger parameter sizes alone do not ensure robustness. Models such as \textit{GPT-4o} and \textit{Gemini-2.5-Pro} exhibit superior contextual adaptability and semantic coherence across both evaluation conditions, sustaining balanced precision and recall in analyzing rule-based contradictions within the \textit{openHAB} dataset. Collectively, these findings underscore that reasoning alignment, rather than scale alone, governs consistent and contextually aware validation of RITs across varying deployment constraints.

\begin{table}[t!]
\centering
\begin{adjustbox}{width=0.95\textwidth}
\begin{tabular}{l|cccc}
\toprule
\textbf{Model} & \textbf{AC} & \textbf{TC} & \textbf{CC} & \textbf{Total Correct} \\
\midrule
Llama-8b Zero-Shot & 92.91\% & 100.00\% & 20.00\% & 88.28\% \\
Llama-8b One-Shot & 88.19\% & 100.00\% & 30.00\% & 84.83\% \\
Llama-8b Two-Shot & 56.69\% & 87.50\% & 30.00\% & 56.55\% \\
\midrule
Llama-70b Zero-Shot & 73.23\% & 50.00\% & 30.00\% & 68.97\% \\
Llama-70b One-Shot & 63.78\% & 50.00\% & 20.00\% & 60.00\% \\
Llama-70b Two-Shot & 66.14\% & 62.50\% & 50.00\% & 64.83\% \\
\midrule
GPT-4o Zero-Shot & 25.20\% & 75.00\% & 90.00\% & 63.40\% \\
GPT-4o One-Shot & 25.20\% & 87.50\% & 80.00\% & 64.23\% \\
GPT-4o Two-Shot & 29.13\% & 87.50\% & 60.00\% & 58.88\% \\
\midrule
Gemini-2.5-Pro Zero-Shot & 57.48\% & 87.50\% & 100.00\% & 62.07\% \\
Gemini-2.5-Pro One-Shot & 55.12\% & 100.00\% & 100.00\% & 60.69\% \\
Gemini-2.5-Pro Two-Shot & 55.91\% & 100.00\% & 100.00\% & 61.38\% \\
\midrule
Deepseek-r1-7b Zero-Shot & 96.06\% & 87.50\% & 30.00\% & 91.03\% \\
Deepseek-r1-7b One-Shot & 96.06\% & 75.00\% & 30.00\% & 90.34\% \\
Deepseek-r1-7b Two-Shot & 55.91\% & 75.00\% & 60.00\% & 57.24\% \\
\bottomrule
\end{tabular}
\end{adjustbox}
\caption{Experiment C: Using only 3 categories of RITs and allowing multiple responses}
\label{table:experiment-c}
\end{table}
\subsection*{Reduced Category Evaluation: Impact of Simplified Threat Taxonomy}
Experiment~C examined model performance when the classification space was reduced to three primary RIT categories---Action Contradiction (AC), Trigger Contradiction (TC), and Condition Contradiction (CC)---while allowing multiple responses. As shown in Table~\ref{table:experiment-c}, simplifying the taxonomy resulted in a substantial accuracy increase across nearly all models, indicating that reducing semantic complexity helps LLMs generalize more effectively across threat types. 

Smaller models such as \textit{Llama-8b} and \textit{Deepseek-r1-7b} achieved the highest overall accuracies (up to 88--91\%) under the zero-shot condition, a marked improvement from their performance under the full six-category setup. This suggests that when the decision space is constrained, these models can more reliably map contextual cues to broad contradiction patterns. Similarly, larger models like \textit{Llama-70b} and \textit{Gemini-2.5-Pro} exhibited stable mid-range accuracies (60--69\%), reflecting improved category alignment with fewer classification boundaries. 

In contrast, \textit{GPT-4o} showed relatively consistent results (58--64\%) across both the three-category and six-category experiments, implying robust generalization regardless of label granularity. This consistency underscores its balanced reasoning and semantic abstraction capability, which prevent overfitting to category structure. 

Comparatively, these findings demonstrate that category simplification narrows the reasoning gap between small and large models, amplifying baseline accuracy but reducing the challenge necessary to differentiate deeper reasoning alignment. In essence, the reduced taxonomy highlights that while structural simplification boosts surface-level accuracy, complex multi-category reasoning remains the true differentiator of advanced, reasoning-optimized architectures such as \textit{GPT-4o} and \textit{Gemini-2.5-Pro}.
\begin{table}[t!]
\centering
\begin{adjustbox}{width=0.95\textwidth}
\begin{tabular}{l|cccc}
\toprule
\textbf{Model} & \textbf{AC} & \textbf{TC} & \textbf{CC} & \textbf{Total Correct} \\
\midrule
Llama-8b Zero-Shot & 52.76\% & 25.00\% & 10.00\% & 48.28\% \\
Llama-8b One-Shot & 58.27\% & 50.00\% & 30.00\% & 55.86\% \\
Llama-8b Two-Shot & 50.39\% & 0.00\% & 0.00\% & 44.14\% \\
\midrule
Llama-70b Zero-Shot & 62.99\% & 12.50\% & 40.00\% & 58.62\% \\
Llama-70b One-Shot & 64.57\% & 25.00\% & 30.00\% & 60.00\% \\
Llama-70b Two-Shot & 67.72\% & 0.00\% & 40.00\% & 62.07\% \\
\midrule
GPT-4o Zero-Shot & 30.71\% & 87.50\% & 60.00\% & 59.40\% \\
GPT-4o One-Shot & 28.35\% & 75.00\% & 60.00\% & 54.45\% \\
GPT-4o Two-Shot & 28.35\% & 75.00\% & 70.00\% & 57.78\% \\
\midrule
Gemini-2.5-Pro Zero-Shot & 48.03\% & 87.50\% & 70.00\% & 51.72\% \\
Gemini-2.5-Pro One-Shot & 44.88\% & 100.00\% & 60.00\% & 48.97\% \\
Gemini-2.5-Pro Two-Shot & 44.88\% & 100.00\% & 60.00\% & 48.97\% \\
\midrule
Deepseek-r1-7b Zero-Shot & 19.69\% & 0.00\% & 40.00\% & 20.00\% \\
Deepseek-r1-7b One-Shot & 21.26\% & 0.00\% & 70.00\% & 23.45\% \\
Deepseek-r1-7b Two-Shot & 22.05\% & 0.00\% & 40.00\% & 22.07\% \\
\bottomrule
\end{tabular}
\end{adjustbox}
\caption{Experiment D: Using only 3 categories of RITs and NOT allowing multiple responses}
v\label{table:experiment-d}
\end{table}

\subsection*{Single-Response Evaluation under Reduced Category Taxonomy}
Experiment~D investigates the impact of enforcing a single-response constraint within the simplified three-category RIT taxonomy (\textit{Action Contradiction (AC)}, \textit{Trigger Contradiction (TC)}, and \textit{Condition Contradiction (CC)}). As shown in Table~\ref{table:experiment-d}, overall performance declined relative to Experiment~C, where multiple responses were permitted. This outcome reinforces the precision--recall trade-off previously observed in Experiment~B, indicating that while reducing the number of categories simplifies the reasoning space, constraining the output to a single label exposes underlying confidence limitations.

Smaller-scale models such as \textit{Llama-8b} and \textit{Deepseek-r1-7b} exhibited marked performance drops (total accuracies below 25--56\%), confirming their sensitivity to the single-label constraint despite benefiting from the reduced taxonomy. Conversely, larger models such as \textit{Llama-70b} and reasoning-oriented architectures like \textit{GPT-4o} maintained moderate stability (58--62\%), demonstrating that scaling and reasoning alignment jointly mitigate overcommitment errors. Interestingly, \textit{Gemini-2.5-Pro} displayed uneven category-specific accuracy—reaching 100\% on TC but with inconsistent total accuracy—suggesting aggressive decision bias toward salient contextual triggers.

Comparatively, while Experiment~C (multi-response) inflated recall and masked uncertainty, Experiment~D offers a clearer depiction of each model’s reasoning discipline under constrained decision pressure. The results confirm that, even in a reduced-category context, alignment-optimized models (\textit{GPT-4o}, \textit{Llama-70b}) sustain more coherent and confidence-calibrated reasoning than smaller or less aligned models. In summary, the interplay between category simplification and single-response enforcement delineates the boundary between surface-level generalization and robust semantic reasoning, extending the findings of RQ1 and RQ2.

\begin{figure}[t!]
    \centering
    \includegraphics[width=0.9\linewidth]{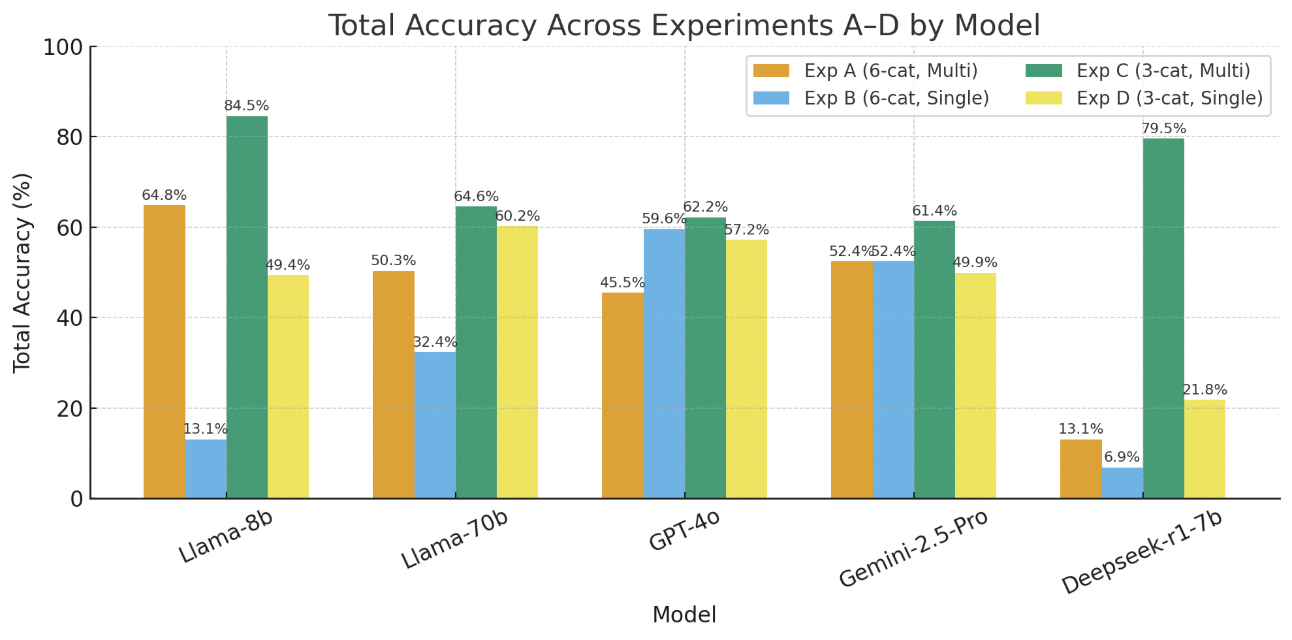}
    \caption{Comparative total accuracy across Experiments A–D for all models, illustrating performance trends under multi- and single-response conditions within six- and three-category RIT taxonomies.}
    \label{fig:expABCD}
\end{figure}

 Overall, accuracy declines consistently across models when constrained to produce a single RIT label, reaffirming the precision–recall trade-off identified in previous analyses. The reduction is most pronounced for smaller architectures such as \textit{Deepseek-r1-7b}, whose total accuracy dropped from over 90\% to approximately 23\%, indicating a sharp confidence loss when forced to commit to a single classification. Conversely, larger or reasoning-aligned models such as \textit{GPT-4o} and \textit{Llama-70b} maintained moderate stability (around 55–60\%), demonstrating more balanced calibration between recall and precision. These findings highlight that while multi-response flexibility benefits recall, single-response evaluation more accurately reflects model reasoning reliability in real-world RIT validation contexts. Figure~\ref{fig:expABCD} visualizes the comparative performance of all evaluated models covering the four experiments (Experiment A-D).

\subsection{Scalability and Generalizability (RQ3)}
\label{sec-mutation}
Building on the earlier experiments with the openHAB and IoTB datasets, RQ3 shifts the focus from accuracy under controlled conditions to robustness at scale. The mutation dataset—orders of magnitude larger and composed entirely of true contradiction cases—presents a far more demanding test of generalization. This setting removes ambiguity, increases structural variety, and exposes models to hundreds of synthetically generated but semantically consistent vulnerabilities
%==================
% Mutation

\begin{table}[t!]
\centering
\begin{adjustbox}{width=\textwidth}
\begin{tabular}{l|ccccccc}
\toprule
\textbf{Model} & \textbf{WAC} & \textbf{SAC} & \textbf{WTC} & \textbf{STC} & \textbf{WCC} & \textbf{SCC} & \textbf{Total Correct} \\
\midrule
Llama-8b Zero-Shot & 74.42\% & 36.44\% & 75.17\% & 22.41\% & 57.78\% & 12.17\% & 42.97\% \\
Llama-8b One-Shot & 60.13\% & 24.47\% & 74.50\% & 12.64\% & 63.72\% & 8.85\% & 39.20\% \\
Llama-8b Two-Shot & 82.06\% & 19.95\% & 83.56\% & 11.49\% & 41.25\% & 4.15\% & 35.19\% \\
\midrule
Llama-70b Zero-Shot & 72.43\% & 9.84\% & 35.91\% & 13.79\% & 53.13\% & 2.07\% & 29.34\% \\
Llama-70b One-Shot & 17.61\% & 4.79\% & 31.21\% & 12.64\% & 29.37\% & 2.77\% & 15.59\% \\
Llama-70b Two-Shot & 37.87\% & 19.68\% & 30.54\% & 20.11\% & 26.48\% & 2.35\% & 19.88\% \\
\midrule
Gemini-2.5-Pro Zero-Shot & 42.86\% & 39.36\% & 87.25\% & 81.03\% & 41.89\% & 51.73\% & 52.63\% \\
Gemini-2.5-Pro One-Shot & 36.21\% & 55.32\% & 76.85\% & 80.46\% & 49.76\% & 63.35\% & 58.28\% \\
Gemini-2.5-Pro Two-Shot & 35.88\% & 60.11\% & 19.13\% & 89.08\% & 48.31\% & 57.95\% & 50.74\% \\
\midrule
Deepseek-r1-7b Zero-Shot & 76.08\% & 36.17\% & 22.15\% & 31.61\% & 30.02\% & 24.07\% & 33.95\% \\
Deepseek-r1-7b One-Shot & 80.40\% & 21.81\% & 8.72\% & 45.40\% & 24.40\% & 24.62\% & 30.42\% \\
Deepseek-r1-7b Two-Shot & 76.08\% & 13.83\% & 11.74\% & 35.06\% & 19.10\% & 25.73\% & 27.33\% \\
\bottomrule
\end{tabular}
\end{adjustbox}
\caption{Experiment A on Mutation Dataset}
\label{tab:experiment_a_mutation}
\end{table}

\section*{Experiment A (6-category taxonomy, multiple responses allowed)}
On the oHC/IoTB dataset, smaller Llama variants and reasoning-optimized models demonstrated mixed strengths. Llama-8b achieved a strong zero-shot total of 64.83\%, performing well on WAC and WCC but poorly on SCC. Gemini-2.5-Pro reached 65.52\% zero-shot, and GPT-4o improved further with two-shot inputs to 60.16\%. Applying the same evaluation to the mutation dataset revealed a notable decline for most models: Llama-8b totals dropped to 42.97\%/39.20\%/35.19\% across 0/1/2-shot settings, while Llama-70b fell from approximately 29.3\% to 15.6\% and then 19.9\%. Deepseek’s performance moved into the mid-30s. In contrast, Gemini-2.5-Pro maintained substantially higher totals on the mutation set (~52.6–58.3\%), indicating stronger generalization to mutation-driven, real-vulnerability instances. Reasoning-aligned architectures retain their advantage under multi-response evaluation, while smaller Llama models degrade substantially when faced with larger, more semantically diverse datasets.  

\begin{table}[t!]
\centering
\begin{adjustbox}{width=\textwidth}
\begin{tabular}{l|ccccccc}
\toprule
\textbf{Model} & \textbf{WAC} & \textbf{SAC} & \textbf{WTC} & \textbf{STC} & \textbf{WCC} & \textbf{SCC} & \textbf{Total Correct} \\
\midrule
Llama-8b Zero-Shot & 15.95\% & 0.80\% & 27.18\% & 37.36\% & 23.27\% & 1.38\% & 14.11\% \\
Llama-8b One-Shot & 11.96\% & 1.06\% & 23.49\% & 13.22\% & 40.61\% & 6.36\% & 17.31\% \\
Llama-8b Two-Shot & 14.95\% & 0.00\% & 24.83\% & 29.31\% & 28.41\% & 0.69\% & 14.11\% \\
\midrule
Llama-70b Zero-Shot & 35.55\% & 6.38\% & 25.50\% & 15.52\% & 38.68\% & 0.00\% & 19.04\% \\
Llama-70b One-Shot & 12.62\% & 7.98\% & 46.98\% & 13.79\% & 37.24\% & 0.28\% & 18.68\% \\
Llama-70b Two-Shot & 16.94\% & 37.77\% & 48.66\% & 23.56\% & 26.48\% & 0.00\% & 21.80\% \\
\midrule
Gemini-2.5-Pro Zero-Shot & 34.55\% & 39.89\% & 80.54\% & 74.14\% & 37.72\% & 48.41\% & 48.42\% \\
Gemini-2.5-Pro One-Shot & 31.89\% & 54.79\% & 72.15\% & 76.44\% & 47.51\% & 60.86\% & 55.55\% \\
Gemini-2.5-Pro Two-Shot & 30.90\% & 58.24\% & 18.79\% & 87.36\% & 46.39\% & 55.46\% & 48.50\% \\
\midrule
Deepseek-r1-7b Zero-Shot & 10.30\% & 4.52\% & 0.67\% & 44.25\% & 6.10\% & 2.35\% & 7.29\% \\
Deepseek-r1-7b One-Shot & 5.65\% & 1.33\% & 0.67\% & 65.52\% & 2.09\% & 4.84\% & 7.45\% \\
Deepseek-r1-7b Two-Shot & 6.31\% & 2.66\% & 0.67\% & 58.05\% & 2.57\% & 10.10\% & 8.86\% \\
\bottomrule
\end{tabular}
\end{adjustbox}
\caption{Experiment B on Mutation Dataset}
\label{tab:experiment_b_mutation}
\end{table}
\section*{Experiment B (6-category taxonomy, single response only)}
Restricting outputs to a single response revealed sharper contrasts. On the oHC/IoTB dataset, single-response totals for many models collapsed (e.g., Llama-8b 13.10\% zero-shot), whereas GPT-4o remained robust (59.6–62.2\% across shots) and Gemini stabilized in the low 50s. On the mutation dataset, single-response performance fell further: Llama-8b remained low (14–17\%), Llama-70b showed modest gains but stayed weak (19–21.8\%), and Deepseek remained ineffective (7–9\%). Gemini-2.5-Pro consistently maintained moderate single-label accuracy across all mutation instances, demonstrating that models optimized for reasoning and alignment are better equipped to handle strict decision constraints. These results indicate that single-response evaluation exposes models’ limitations in confidence calibration, and reasoning/alignment capabilities enhance accuracy in selecting a correct, decisive label.  

\begin{table}[t!]
\centering
\begin{adjustbox}{width=0.95\textwidth}
\begin{tabular}{l|cccc}
\toprule
\textbf{Model} & \textbf{AC} & \textbf{TC} & \textbf{CC} & \textbf{Total Correct} \\
\midrule
Llama-8b Zero-Shot & 84.93\% & 96.82\% & 27.19\% & 56.03\% \\
Llama-8b One-Shot & 77.55\% & 94.70\% & 20.51\% & 50.02\% \\
Llama-8b Two-Shot & 55.39\% & 70.97\% & 48.51\% & 54.63\% \\
\midrule
Llama-70b Zero-Shot & 79.76\% & 40.04\% & 31.72\% & 46.33\% \\
Llama-70b One-Shot & 75.63\% & 23.73\% & 23.40\% & 37.64\% \\
Llama-70b Two-Shot & 65.58\% & 34.53\% & 41.75\% & 46.85\% \\
\midrule
Gemini-2.5-Pro Zero-Shot & 59.53\% & 94.28\% & 99.55\% & 87.70\% \\
Gemini-2.5-Pro One-Shot & 57.02\% & 93.43\% & 98.14\% & 86.09\% \\
Gemini-2.5-Pro Two-Shot & 49.93\% & 93.86\% & 98.22\% & 84.29\% \\
\midrule
Deepseek-r1-7b Zero-Shot & 94.68\% & 83.26\% & 30.39\% & 57.84\% \\
Deepseek-r1-7b One-Shot & 97.64\% & 83.69\% & 28.53\% & 57.72\% \\
Deepseek-r1-7b Two-Shot & 66.47\% & 67.16\% & 68.57\% & 67.74\% \\
\bottomrule
\end{tabular}
\end{adjustbox}
\caption{Experiment C on Mutation Dataset}
\label{tab:experiment_c_mutation}
\end{table}

\section*{Experiment C (3-category taxonomy, multiple responses allowed)}
Reducing the label granularity led to improved absolute totals across all models, a pattern that persisted on the mutation dataset. In mutation Experiment C, Gemini-2.5-Pro achieved very high totals (\textasciitilde84–88\% across shots), and Deepseek showed notable improvement (\textasciitilde57–67\%). Llama-8b and Llama-70b improved modestly (\textasciitilde50–56\% and \textasciitilde37–46\%, respectively) but continued to lag behind reasoning-optimized models. Although the mutation dataset narrowed the gap for smaller models in multi-response mode, reasoning-aligned architectures continued to capture the largest gains. Reducing taxonomy benefits all models, but alignment and reasoning optimization provide the most robust and generalizable performance.  

\begin{table}[t!]
\centering
\begin{adjustbox}{width=0.95\textwidth}
\begin{tabular}{l|cccc}
\toprule
\textbf{Model} & \textbf{AC} & \textbf{TC} & \textbf{CC} & \textbf{Total Correct} \\
\midrule
Llama-8b Zero-Shot & 41.95\% & 30.30\% & 23.70\% & 29.90\% \\
Llama-8b One-Shot & 44.61\% & 23.31\% & 27.12\% & 31.14\% \\
Llama-8b Two-Shot & 53.47\% & 31.99\% & 22.88\% & 32.91\% \\
\midrule
Llama-70b Zero-Shot & 63.81\% & 23.94\% & 35.51\% & 41.00\% \\
Llama-70b One-Shot & 53.32\% & 23.94\% & 33.58\% & 37.11\% \\
Llama-70b Two-Shot & 62.48\% & 20.13\% & 26.97\% & 35.31\% \\
\midrule
Gemini-2.5-Pro Zero-Shot & 46.53\% & 92.37\% & 73.11\% & 69.54\% \\
Gemini-2.5-Pro One-Shot & 45.49\% & 91.95\% & 75.63\% & 70.54\% \\
Gemini-2.5-Pro Two-Shot & 44.46\% & 79.24\% & 80.24\% & 70.34\% \\
\midrule
Deepseek-r1-7b Zero-Shot & 34.42\% & 6.78\% & 35.29\% & 29.66\% \\
Deepseek-r1-7b One-Shot & 34.12\% & 8.26\% & 44.73\% & 34.95\% \\
Deepseek-r1-7b Two-Shot & 26.14\% & 9.96\% & 47.18\% & 34.43\% \\
\bottomrule
\end{tabular}
\end{adjustbox}
\caption{Experiment D on Mutation Dataset}
\label{tab:experiment_d_mutation}
\end{table}
\begin{figure}[t!]
    \centering
    \includegraphics[width=0.99\linewidth]{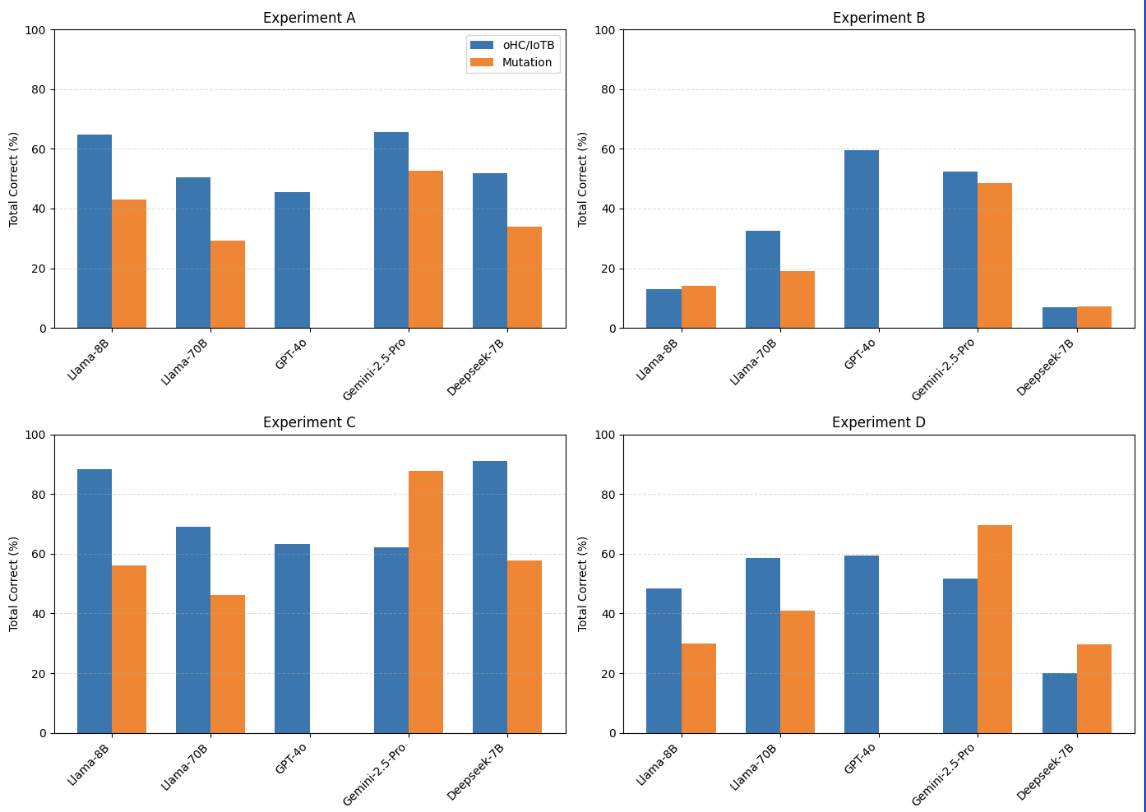}
    \caption{Comparative total accuracy across Experiments A–D (Mutation Vs oHC/IoTB dataset). for all models, illustrating performance trends under multi- and single-response conditions within six- and three-category RIT taxonomies.}
    \label{fig:expABCDM}
\end{figure}
\section*{Experiment D (3-category taxonomy, single response only)}
When the three-category space was combined with a single-response constraint, overall performance was lower than multi-response but higher than six-category single-response evaluations for some models. On the mutation dataset, Gemini-2.5-Pro maintained strong single-response totals (69.5–70.5\%), indicating high precision and stability under stricter decision conditions. Llama-70b scored in the mid-30s to low-40s (35–41\%), Deepseek hovered in the low-30s, and Llama-8b improved slightly compared to six-category single-response performance (29–33\%). Although the mutation dataset reduced some advantages observed on oHC/IoTB, reasoning-aligned models preserved single-label capability far better than non-aligned Llama variants, demonstrating both consistency and resilience under constrained evaluation.  
%\section*{Synthesis for RQ3 (Scalability and Generalizability)}

Across all four experiments, a clear pattern emerges. Models with explicit reasoning and alignment, such as Gemini-2.5-Pro and, to a lesser extent, GPT-4o, maintain performance advantages on large mutation datasets and under stricter single-response constraints. These models generalize more effectively, even when every instance represents a true vulnerability. In contrast, Llama-family models, especially larger variants without alignment tuning, and smaller architectures show substantial performance degradation on mutation data, particularly under single-response or multi-category pressure. Reducing the taxonomy to three categories improves accuracy for all models but masks deeper differences in generalization. Alignment-optimized models retain higher absolute performance and robust generalization, whereas non-aligned models gain only modestly and remain brittle. Collectively, these findings indicate that true scalability and generalization are achieved primarily through reasoning and alignment capabilities rather than parameter count alone.  
\subsection{Comparison to Static Analysis}
\hlc[highlight]{As a follow-up experiment, LLMs were then employed in a hybrid method with an existing static analysis tool. The static analysis tool oHIT was designed to detect RITs in openHAB systems. The tool demonstrated exceptional recall in detecting threats, but was weaker in overall precision, ie it rarely produces false negatives, but, like many static analysis tools, is prone to producing false positives. With this in mind, we employed the LLMs as an additional layer on top of oHIT, using them to adjudicate whether the threats detected by static analysis were true or false positives.
This is done in a multi-step process, noticing the specific weaknesses of oHIT, and then breaking the problem down into the simplest possible steps for the LLM to adjudicate upon. oHIT has weak precision in determining the trigger cascade threats. This is because they are so contextual. A trigger cascade can be a problematic error in the code, or, more commonly, it is completely intentional by the user}  

While LLMs showed promise in detecting RITs, their performance fell short of the recall and consistency achieved by static analysis tools like oHIT. The models’ inability to reliably differentiate between similar RIT patterns and their dependence on prompt design underscore the need for hybrid approaches that combine symbolic reasoning with LLMs to reduce false positives and improve accuracy.
\begin{table}[t!]
\centering
\begin{adjustbox}{width=0.99\textwidth}
\begin{tabular}{lc|cccccc|c}
  \toprule
  \textbf{Tool} &
  \textbf{Method} &
  \textbf{WAC} &
  \textbf{SAC} &
  \textbf{WTC} &
  \textbf{STC} &
  \textbf{WCC} &
  \textbf{SCC} &
  \textbf{Total} \\
  \midrule
  oHIT & Static                      & 83.68\%  & 88.24\%  & 16.67\%  & 46.15\%  & 100.00\% & 100.00\% & 72.53\%  \\
  \midrule
  oHIT + Llama-8b & Zero-Shot        & 62.24\% & 41.18\% & 50.00\% & 76.92\% & 100.00\% & 100.00\% &  71.72\%   \\
  oHIT + Llama-8b & One-Shot         & 29.59\% & 58.82\% & 66.67\% & 53.85\% & 100.00\% & 100.00\% &  68.15\%   \\
  oHIT + Llama-8b & Two-Shot         & 33.67\% & 45.10\% & 66.67\% & 61.54\% & 100.00\% & 100.00\% &  67.83\%   \\
  \midrule
  oHIT + GPT-4o & Zero-Shot          & 91.84\% & 94.12\% & 58.33\% & 69.23\% & 100.00\% & 100.00\% &  85.59\% \\
  oHIT + GPT-4o & One-Shot           & 90.82\% & 96.08\% & 41.67\% & 53.85\% & 100.00\% & 100.00\% &  80.40\%   \\
  oHIT + GPT-4o & Two-Shot           & 92.86\% & 98.04\% & 66.67\% & 76.92\% & 100.00\% & 100.00\% &  89.08\%   \\
  \midrule
  oHIT + Gemini 2.5 & Zero-Shot      & 89.80\% & 96.08\% & 83.33\% & 53.85\% & 100.00\% & 100.00\% &  87.18\%   \\
  oHIT + Gemini 2.5 & One-Shot       & 89.80\% & 98.04\% & 83.33\% & 84.62\% & 100.00\% & 100.00\% &  92.63\%   \\
  oHIT + Gemini 2.5 & Two-Shot       & 89.80\% & 98.04\% & 83.33\% & 84.62\% & 100.00\% & 100.00\% &  92.63\%   \\
  
  \bottomrule
\end{tabular}
\end{adjustbox}
\caption{Experiment G: Complementing static analysis tool by detecting false positives from oHIT}
\label{table:experiment-g-transposed}
\end{table}

\subsection{Hybrid Approach: Combining LLMs with Static Analysis}
This subsection answers \textbf{RQ4}. In static analysis, we noticed that false positives occured for a finite, identifiable set of reasons, typically related to the static analysis tool's lack of contextual understanding. For example, one of the main sources of false positives in oHIT arises during the determination of trigger overlap. For example, oHIT might incorrectly determine that triggers overlap due to variable naming conventions, such as when Trigger A is set to activate at \textit{"cron '0 8 * * *'"} (8am) and Trigger B is activated when \textit{"Sun\_Is\_Setting\_Event changes to ON"}. While it is clear to a human that 8am and sunset will never overlap, static analysis lacks the common sense to make this distinction. 

Another significant source of false positives in oHIT is its detection of trigger cascade threats. Trigger cascades can be either dangerous or intentional, and static analysis cannot distinguish between the two.

To assess how LLMs can augment the symbolic rule-interaction pipeline, we evaluated three hybrid configurations — oHIT + Llama-8B, oHIT + GPT-4o, and oHIT + Gemini 2.5 (Reasoning) — on the oHC + IoTB dataset. These experiments reuse oHIT’s static detections as inputs to an LLM “adjudication” stage, where the model determines whether each candidate interaction constitutes a true or false RIT. The reported “Total” values in Table \ref{table:experiment-g-transposed} represents the final precision that was achieved after the LLM adjudication step.
To control runtime cost while representing distinct reasoning styles, we selected three complementary models:

\begin{itemize}
    \item{Llama-8B – a lightweight, cost-efficient non-reasoning baseline;} 
    \item{GPT-4o – a large, high-fidelity non-reasoning model;}
    \item{Gemini 2.5 Reasoning – a reasoning-oriented model replacing DeepSeek-R1 after pilot trials showed higher consistency.}
\end{itemize}

Each hybrid variant was run under zero-, one-, and two-shot prompt configurations using the following generation parameters: temperature = 0.2, top\_p = 0.95, max\_output\_tokens = 2048.

The adjudication pipeline decomposed oHIT’s RIT candidates into fine-grained reasoning subtasks, each answered independently by the LLM:
\begin{itemize}
\item Trigger-Overlap Analysis – Determines whether two rule triggers can activate simultaneously (e.g., sunrise\_event vs. sunset\_event).

\item Trigger-Cascade Safety – Judges whether a trigger cascade was likely intended by the user or if it potentially produces undesirable behaviour.

\item Action-Conflict Check – Compares two actions to decide if they assign incompatible values to the same device attribute.
\end{itemize}

\begin{table}[t!]
\centering
\begin{adjustbox}{width=0.95\textwidth}
%\label{tab:ohit_mutation}
\centering
\begin{tabular}{|l|c|c|c|c|}
\hline
\textbf{Category} & \textbf{Total Rulesets} & \textbf{Correctly Detected} & \textbf{Undetected} & \textbf{Detection Rate} \\
\hline
SAC & 376 & 376 & 0 & 100.00\% \\
WAC & 301 & 299 & 2 & 99.34\% \\
STC & 174 & 158 & 16 & 90.80\% \\
WTC & 298 & 263 & 35 & 88.26\% \\
SCC & 723 & 690 & 33 & 95.44\% \\
WCC & 623 & 595 & 28 & 95.51\% \\
\hline
\textbf{Total} & \textbf{2,495} & \textbf{2,381} & \textbf{114} & \textbf{95.43\%} \\
\hline
\end{tabular}
\end{adjustbox}
\caption{oHIT Performance on Mutation Dataset}
\label{tab:ohit_mutation}
\end{table}

\begin{table}[t!]
\centering
\begin{adjustbox}{width=0.95\textwidth}
%\label{tab:gemini_ohit_fn}
\centering
\begin{tabular}{|l|c|c|c|c|}
\hline
\textbf{Category} & \textbf{Total Rulesets} & \textbf{Correctly Detected} & \textbf{Undetected} & \textbf{Detection Rate} \\
\hline
SAC & 0 & - & - & - \\
WAC & 2 & 2 & 0 & 100.00\% \\
STC & 16 & 2 & 14 & 12.50\% \\
WTC & 35 & 18 & 17 & 51.43\% \\
SCC & 33 & 33 & 0 & 100.00\% \\
WCC & 28 & 6 & 22 & 21.43\% \\
\hline
\textbf{Total} & \textbf{114} & \textbf{61} & \textbf{53} & \textbf{53.51\%} \\
\hline
\end{tabular}
\end{adjustbox}
\caption{Top model performance on false negatives produced by oHIT on the mutation dataset}
\label{tab:gemini_ohit_fn}
\end{table}
The hybrid architecture consistently reduced false positives by leveraging LLM semantic priors about trigger intent and device relationships. The best results were produced with the Gemini 2.5 model. Precision improved or remained stable for every RIT class, increasing the overall precision increased from 0.73 (oHIT-only baseline) to 0.93 under the best hybrid setting (two-shot). These results confirm that a modular symbolic + LLM pipeline can meaningfully improve IoT safety verification accuracy without retraining or altering the original static analysis engine.

In addition to adding an LLM to the static analysis pipeline to adjudicate on true/false positives, we mirrored the experiment to determine its value in adjudicating true/false negatives. As oHIT achieved 100\% recall on the combined IotB and oHC datasets, the false negatives were derived from oHIT's performance on the mutation dataset. The prompting strategy for this recovery phase differs fundamentally from the validation phase. When determining if a positive identification is valid, the LLM is provided with context clues from the static tool. In particular,, which RIT was flagged, the relevant conditions, and the specific rules involved. However, when feeding the LLM negatives, these context clues are absent; the LLM must perform a blind search to identify threats that the static analysis missed.

Table \ref{tab:ohit_mutation} details the performance of oHIT on the mutation dataset. While the tool performs exceptionally well with a 95.43\% detection rate, it produces misses stemming from two primary sources: type handling discrepancies and pipeline rigidity. Specifically, the condition cascade mutations (SCC/WCC) occasionally produce complex types (e.g., PointType) that the static parser fails to resolve, while trigger cascade mutations (STC/WTC) relying on openHAB’s internal state propagation (postUpdate triggering sendCommand) are rejected by oHIT’s stricter event matching logic.

The results of the LLM’s attempt to recover these missed threats are presented in Table \ref{tab:gemini_ohit_fn}.  For STC and WTC, the results suggest that the specific intricacies causing static analysis failures, largely to do with unmatching discrepancies between posting an update and sending a command, are equally opaque to the LLM. Conversely, the model performed extremely well on SCC, successfully recovering 100\% of the false negatives. This success highlights a key complementarity: oHIT missed these threats because it was not robust to the syntactical anomalies produced by the mutation tool, whereas the LLM could adapt to these irregularities through semantic approximation. The WCC results closely match the model’s general performance in Experiment A, confirming that "weak" threats involving partial conditions remain a consistent challenge for the model regardless of the experimental setup.

\subsection{Token Lengths, Context Usage, and Performance Characteristics}
Across all experiments, the combined system prompts, instructions, and rule sets produced an upper prompt length of ~2,200 tokens. The largest base prompt—Experiment B (two-shot)—contained 1,992 tokens, computed using each model’s default tokenizer from its corresponding .gguf file. With an 8,192-token context window configured for all local models, no input truncation occurred, Table \ref{tab:tokens_per_prompt}.

Performance analysis revealed clear distinctions between deployment modalities. The Llama 3.1 8B model achieved rapid local inference (~2.57 s) and remained within the memory capabilities of modern edge accelerators. DeepSeek-R1 7B, despite its efficient 5.3 GB footprint, incurred significantly higher latency (~20.42 s) due to its reasoning pipeline. Cloud-based inference removed hardware constraints but introduced substantial delay; Gemini-2.5-Pro averaged 24.55 s per response, Table \ref{tab:generation_times}.

Reliability differences were pronounced. Gemini produced one blank output and accumulated 6,340 total HTTP errors, HTTP 429 (Too Many Requests) and HTTP 503 (Service Unavailable) errors, occasionally requiring multiple API attempts to obtain a valid response. These failures highlight a structural vulnerability in cloud-offloaded control architectures, where API quotas and service availability undermine real-time guarantees.

Overall, while cloud LRMs offer stronger reasoning capabilities, local execution remains essential for deterministic, latency-sensitive IoT tasks; a hybrid approach may be required for robust autonomous operation.
\begin{table}[t!]
\centering
\begin{adjustbox}{width=0.7\textwidth}
\begin{tabular}{l|cc|cc}
\toprule
\textbf{RIT Classification} & \multicolumn{2}{c|}{\textbf{3-Letter RITs}} & \multicolumn{2}{c}{\textbf{2-Letter RITs}} \\
\midrule
\textbf{Experiment} & \textbf{Exp A} & \textbf{Exp B} & \textbf{Exp C} & \textbf{Exp D} \\
\midrule
\textbf{0-Shot} & 492 & 527 & 441 & 480 \\
\textbf{1-Shot} & 961 & 997 & 703 & 753 \\
\textbf{2-Shot} & 1928 & 1992 & 935 & 972 \\
\bottomrule
\end{tabular}
\end{adjustbox}
\caption{Token Count per Prompt by Experiment and Prompting Strategy (Using Ollama Default Tokenizer, Excluding Model Rulesets)}
\label{tab:tokens_per_prompt}
\end{table}

\begin{table}[t!]
\centering
\begin{adjustbox}{width=0.99\textwidth}
\begin{tabular}{l|ccc|c}
\toprule
\textbf{Compute Resource} & \multicolumn{3}{c|}{\textbf{Mobile 3070ti 8GB GDDR6}} & \textbf{H100 Server} \\
\midrule
\textbf{Models} & \textbf{Llama 3.1 8b} & \textbf{Deepseek R1 7b} & \textbf{Gemini-2.5-Pro} & \textbf{Llama 3.1 70b} \\
\midrule
\textbf{Avg Generation Time} & 2.57s & 20.42s & 24.55s & 13.58s \\
\bottomrule
\end{tabular}
\end{adjustbox}
\caption{Average Generation Times by Model and Compute Resource}
\label{tab:generation_times}
\end{table}
\section{Related Work}\label{sec5}
\subsection{IoT Vulnerability Detection Tools}
There are many proposed methods for analyzing IoT apps to ensure safety. Notable tools include the following: Soteria \cite{Soteria}, IoTSan \cite{Iotsan}, iRuler \cite{iruler}, and IoTCom \cite{iotcomm} all use static analysis techniques, finding some intermediate representation of the source code and feeding it into a model checker to analyze device interactions.

Key tools in IoT analysis include Soteria \cite{Soteria}, IoTSan \cite{Iotsan}, iRuler \cite{iruler}, and IoTCom \cite{iotcomm}. These tools use static analysis by extracting an intermediate code representation and applying model checking to examine SmartThings and IFTTT applications. Other tools, such as IoTGuard \cite{iotguard}, IoTBox \cite{iotbox}, and HaWatcher \cite{HAWatcher}, rely on dynamic methods, analyzing programs during execution \cite{Survey}. Hybrid approaches combine both methods, as seen in SmartFuzz \cite{SmartFuzz} and IoTSAFE \cite{Iotsafe} \cite{Survey}. Tools like SafeTAP \cite{Safetap}, AutoTAP \cite{Autotap}, and MenShen \cite{MenShen} focus on IoT system design, helping users create safe device rules from the outset. These advancements have enabled deeper analysis of device interactions, event flows, and permissions in SmartThings apps, improving safety and reliability.

In open-source home automation, openHAB is a widely used platform, but safety-focused development remains limited. Two tools aim to address multiple platforms. VISCR \cite{viscr} generates tree-based code abstractions and uses graph-based policies for platforms like SmartThings, openHAB, IFTTT, and MUD apps. PatrIoT \cite{Patriot} specifies safety properties and prevents policy violations during system operation, supporting SmartThings, openHAB, and EVA ICS.
\subsection{LLM Approaches to Software Vulnerability Detection}
A review of recent literature reveals a significant trend in leveraging LLMs for source code safety analysis, with varied approaches and objectives. Several studies explore the direct application of LLMs, such as GPT-4 and Claude, for vulnerability detection in specific contexts like smart contracts.  For example, Chen et al. assess ChatGPT's effectiveness in detecting smart contract vulnerabilities using prompt engineering and compare it with other detection tools \cite{chen2024}.  David et al. focus on GPT-4 and Claude by using a chain-of-thought approach to audit smart contracts, and examine how the length of code contexts affect their analyses \cite{david2023}. These studies show that while LLMs can be effective at identifying vulnerabilities they can also produce false positives and have limitations with long code contexts.  

Other research explores the capabilities of more specialised LLMs, like CodeBERT and models in the GPT family for different kinds of analysis such as code completion and bug fixing. For example, Omar et al. employ a transformer-based language model based on CodeBERT and fine-tune it for software vulnerability detection and compare it to LSTM models and static analysis tools \cite{omar2023}. Pearce et al. investigate zero-shot vulnerability repair using Codex, Jurassic-1, and polycoder and different prompting techniques \cite{pearce2022}.  Sandoval et al. assess the security impacts of a Codex powered code completion assistant on code produced by developers \cite{sandoval2023}.

Many studies focus on improving the performance of LLMs, often through specialized training or techniques. For instance, in \cite{ferrag2024} Ferrag et al. introduce SecureFalcon, a fine-tuned lightweight LLM, that uses transfer learning and contrastive learning, while He et al. explore controlled code generation using Codex, PaLM, AlphaCode, and CodeGen \cite{He_2023}. Setak et al. look at using a teacher LLM to improve the ability of an LLM to perform code mutation, with experiments on Llama3 \cite{setak2024}, and Tihanyi et al. propose an approach where an LLM is integrated with the ESBMC model checker, taking the model checker's output as prompts when repairing code \cite{tihanyi2024}. Venkatesh et al. provide a broad evaluation of 26 LLMs including GPT 3.5 Turbo, GPT 4, and llama2, for call graph analysis and type inference of Python programs \cite{venkatesh2024}. Finally, Zhou et al. compare 12 open source LLMs and ChatGPT with 15 SAST tools, and introduce a novel approach for repo-level vulnerability detection \cite{zhou2024}. They use various prompt-based methods like zero-shot prompting, chain-of-thought (CoT) prompting and fine-tuning. These papers demonstrate the potential of fine-tuning, specialised training, and other approaches to optimise LLMs for source code analysis tasks.

To contextualize our investigation into Rule Interaction Threats (RITs), we examined recent comprehensive surveys that map the landscape of LLM applications in software security and the specific challenges of detecting IoT vulnerabilities. These works provide the theoretical foundation for our hybrid approach, highlighting both the potential of generative models and the persistent necessity of static analysis in safety-critical environments.

Zhu et al. \cite{ZhuSurvey} provided a systematic review of Large Language Models (LLMs) in software security, categorizing their application into key areas such as fuzzing, unit testing, program repair, and bug triage. Their analysis breaks down these techniques into stages—pre-processing, prompt generation, and post-processing—observing that while LLMs significantly reduce manual effort, they still face significant limitations. An important insight from their work, which mirrors our experimental results in RQ1, is that out-of-the-box LLMs operating in zero-shot settings often struggle with precise vulnerability detection, yielding low accuracy that motivates the need for fine-tuning or more guided use. This limitation supports our decision to implement the "LLM-oHIT Reconciliation" workflow, where the LLM serves as a post-processing validator rather than a standalone detection engine. 

Conversely, Zhou et al. \cite{ZhouSurvey} examined the security implications inherent to the models themselves, focusing on risks such as hallucination, bias, and reliability. Their survey highlights that while models like ChatGPT demonstrate significant improvements in Natural Language Processing (NLP), they are prone to generating plausible yet false content—a phenomenon known as hallucination—and are highly sensitive to adversarial manipulation. This analysis provides the theoretical basis for the performance degradation we observed in the Mutation dataset (RQ3). When faced with the strict structural logic required for Strong Trigger Cascades (STC), the models often prioritized semantic fluency over logical correctness, a reliability gap that Zhou et al. identify as a key barrier to deploying LLMs in high-stakes decision-making applications. Vulnerability Detection in IoT Firmware Narrowing the scope to the Internet of Things (IoT), 

Feng et al. \cite{FengSurvey} classified vulnerability detection methodologies into four distinct categories: emulator-based testing, automatic code analysis, network fuzzing, and manual reverse engineering. Their survey highlights the unique challenges of the IoT domain, specifically the diversity of architectures and the difficulty of creating accurate emulation environments for dynamic analysis. Most importantly, Feng et al. identified that while automatic code analysis (static analysis), our baseline oHIT tool's category, can work without device entities, and that code-matching solutions like this are suitable for large-scale testing. They also emphasize that such approaches inevitably face issues with false positives and false negatives, making it challenging to strike a balance between efficiency and accuracy. In our work, we interpret these errors as symptoms of static analysis lacking runtime context. Our research directly addresses this specific gap identified by Feng et al. by proposing a hybrid architecture that leverages the semantic understanding of LLMs to filter the context-blind false positives generated by traditional static analysis tools.

\section{Threats to Validity}
\textbf{Internal Validity.} This study employs both naturally occurring openHAB rules and systematically generated mutation rules, providing a broad range of interaction patterns. Although the original oHC/IoTB dataset is imbalanced---with Weak Action Contradictions (WACs) representing a larger portion of the ground truth---this distribution reflects real-world deployments and therefore strengthens ecological validity. To ensure that no category dominated the evaluation, we applied normalization and conducted four controlled experiments (A--D) to isolate the effects of category scope and response constraints. While LLMs may generate multiple plausible RITs, evaluation was standardized against oHIT's verified labels to maintain methodological consistency. This conservative scoring provides a rigorous lower bound on LLM performance.

\textbf{Construct Validity.}  
The threat categories (WAC, SAC, WTC, STC, WCC, SCC) are derived from established TAC-based taxonomies and were applied consistently across symbolic and LLM-based analyses. By evaluating models under four controlled experimental configurations, we explicitly examined whether model behavior depends on category granularity or output constraints. Although LLMs sometimes propose alternative threats beyond oHIT’s classification, our scoring framework prioritizes reproducibility and precision. Future work may extend evaluation to include validation of additional LLM-discovered threats, but the present approach ensures stable constructs and comparability across all models and datasets.

\textbf{External Validity.}  
The use of openHAB rules from two independent community repositories, combined with a large mutation-based corpus, enhances the coverage of syntactic and semantic variations. The mutation dataset introduces rule structures that are infrequent in natural repositories yet valuable for assessing LLM generalization. While findings may not directly extend to all IoT platforms, the TAC model used in openHAB is representative of contemporary smart home ecosystems. Moreover, inclusion of multiple model families (Llama~3.1, GPT-4o, Gemini~2.5~Pro, DeepSeek-r1) supports generalizability across architectural paradigms and parameter scales.

\textbf{Conclusion Validity.}  
Model performance was measured using per-category accuracy as well as aggregated total-correct metrics across four controlled settings. These complementary evaluation perspectives help mitigate bias caused by dataset imbalance or model overfitting to dominant categories. The dual evaluation across natural and mutation-based datasets strengthens reliability, particularly in assessing model robustness under structural variation. 

To summarize,  the combination of natural and synthetic datasets, multiple prompting regimes, and controlled evaluation configurations provides a methodologically rigorous foundation for the reported findings. While opportunities remain for broader dataset balancing and targeted model fine-tuning, the current experimental design offers a strong and conservative assessment of LLM capabilities in detecting rule interaction threats.

\section{Conclusion}\label{sec7}
%In this paper, we evaluated the capability of Large Language Models (LLMs), specifically LlaMa 3.1 8B and GPT-4o, in detecting RITs in openHAB automation rules, comparing their performance to a symbolic reasoning-based static analysis tool, oHIT. Our experiments revealed that while LLMs demonstrate potential in identifying certain RITs, their performance varies significantly with threat complexity. LlaMa 3.1 8B struggled with fine-grained distinctions between weak and strong RITs, while GPT-4o excelled in simpler patterns like Strong Trigger Cascades but underperformed in more complex scenarios such as Weak Action Contradictions. Symbolic reasoning, on the other hand, provided more consistent and precise results, particularly for deeply nested threats.

%A key finding of this work is the potential of a hybrid approach, combining symbolic reasoning with LLMs, to reduce false positives and improve precision. By leveraging the common sense and contextual understanding of LLMs to validate trigger overlaps and intentionality in rule interactions, we significantly enhanced oHIT’s precision—from 84\% to 93\% for WAC threats and from 17\% to 67\% for WTC threats. 

%Future work could explore further optimization of hybrid approaches, incorporating larger LLMs with enhanced reasoning capabilities, and expanding the dataset to include more diverse rule interaction patterns. By continuing to refine these methodologies, we can move closer to achieving reliable and scalable solutions for detecting and mitigating RITs in IoT systems.
This study provides a comprehensive evaluation of LLM-based interaction threat detection across diverse rule representations, threat categories, and prompt configurations. Our results highlight a clear pattern: while modern LLMs demonstrate meaningful semantic understanding of \emph{Trigger--Action--Condition} rules—particularly for threats involving recognizable action and condition relationships—their reliability deteriorates sharply when deeper structural reasoning is required. This limitation becomes most pronounced in the Mutation dataset. Even the strongest models exhibit sizable performance drops, underscoring that current LLMs are highly sensitive to surface form and lack the robustness needed for dependable automation-safety analysis.
Across experiments, no single model or prompting strategy delivered stable performance, and gains in one threat category often coincided with regressions in others. In contrast, the symbolic reasoning baseline remained consistent across both datasets, unaffected by structural rewrites and better aligned with the needs of safety-critical IoT environments. These findings suggest that LLMs, in their current form, cannot replace formal or symbolic analyses for interaction-threat detection, particularly when rules exhibit complex interdependencies or nontrivial structural transformations.
However, the models’ strengths in semantic pattern recognition point toward a promising path forward. A hybrid approach—pairing symbolic reasoning for structural guarantees with LLMs for semantic interpretation and ambiguity resolution—offers a balanced foundation for next-generation analysis tools. Such architectures can leverage LLMs to reduce noise and enrich context, while relying on symbolic methods to ensure correctness, robustness, and reproducibility.
Overall, this work demonstrates both the opportunities and the current limitations of applying LLMs to IoT interaction-threat detection. It establishes a clear direction for future research: building principled hybrid systems that combine the complementary strengths of symbolic verification and large-scale language understanding to provide reliable security analysis for smart home ecosystems.

\section*{Declarations}
\begin{itemize}
\item Funding: Not applicable
\item Ethical approval: Not applicable
\item Informed consent: Yes
\item Author Contributions: Experiments and writing performed by first \& second authors Noura Khajehnouri \& Jason Quantrill, with experiment design, support, guidance, and editing by third author Dr. Manar Alalfi
\item Data Availability Statement: The data pertaining to this research can be found at \url{https://github.com/JasonQuantrill/llm-v-static-results}
\item Conflict of Interest: Not applicable
\item Clinical Trial Number in the manuscript: Not applicable
\end{itemize}

%\nocite*
\bibliographystyle{plain}      % basic style, author-year citations
\bibliography{LLMbib}% common bib file
\end{document}